\documentclass[12pt,cite,a4paper]{article}

\pdfoutput=1

\usepackage{amsmath}
\usepackage{amssymb}
\usepackage{a4wide}
\usepackage{graphicx}

\numberwithin{equation}{section}

\begin{document}

\rightline{IFUP-TH 2008/11}
\vspace{2.2truecm}

%%%%%%%%%%%%%%%%%
\centerline{\Large \bf  Giant magnons and spiky strings on the conifold}

\vspace{1.5truecm}

\centerline{
    {\large Sergio Benvenuti$^{a,}$}\footnote{sbenvenu@princeton.edu}
    {\large \,and\,}
    {\large Erik Tonni${}^{b,}$}\footnote{tonni@df.unipi.it}}

\vspace{1.5cm}

\centerline{{\it ${}^a$Department of Physics, 
Princeton University}}
\centerline{{\it Princeton, NJ 08544, USA}}
\vspace{.5cm}
\centerline{{\it ${}^b$ Dipartmento di Fisica, 
Universit\`a di Pisa and INFN, sezione di Pisa,}}  
\centerline{{\it Largo Bruno Pontecorvo 3, 56127 Pisa, Italy}}

\vspace{2.5truecm}

%%%%%%%%%%%%%%%%%
\centerline{\bf Abstract}
\vspace{.5truecm}

We find explicit solutions for giant magnons and spiky strings on the squashed three dimensional sphere. For a special value of the squashing parameter the solutions describe strings moving in a sector of the conifold, while for another value of the squashing parameter we recover the known results on the round three dimensional sphere. A new feature is that the energy and the momenta enter in the dispersion relation of the conifold in a transcendental way.

\vspace{2truecm}

%%%%%%%%%%%%%%%%%%%%%%%%%%%%%%%%%%%%%%%%%%%%%%%%%%
\newpage

\section{Introduction}

Integrability is the key concept underlying the recent advances in understanding the AdS/CFT correspondence in the planar limit. On the gauge theory side, the dilatation operator acts on the long single trace operators  as the Hamiltonian of a certain spin chain on the states of its Hilbert space; therefore the powerful method of the algebraic Bethe ansatz can be employed to diagonalize the matrix of the anomalous dimensions \cite{Minahan:2002ve}. The insertions of different operators in the trace are treated as impurities (magnons) in the spin chain and integrability tells that a generic scattering process factorizes into two-magnons scatterings. This means that the dispersion relation for a single magnon  and the two-magnon $S$ matrix are enough to determine all the spectrum. 
For the main example of AdS/CFT, namely the correspondence between type IIB string theory on $AdS_5 \times S^5$ and  $\mathcal{N}=4$ SYM, much progress has been done in this direction and now both the dispersion relation and the $S$ matrix are available \cite{Beisert:2005tm, Beisert:2006ez, Beisert:2006ib, Staudacher:2004tk, Beisert:2005fw, Arutyunov:2004vx, Janik:2006dc, Bena:2003wd}.

On the string side, the classical solutions corresponding to one magnon excitation are called giant magnons: they are certain open strings moving in the compact part of the background with finite angular amplitude. Remarkably, the dispersion relation among their conserved charges gives the strong coupling limit of the one found from the gauge theory analysis \cite{Hofman:2006xt, Dorey:2006dq, Chen:2006gea, Arutyunov:2006gs, Astolfi:2007uz, Spradlin:2006wk, Kruczenski:2006pk}. The finite angular amplitude of the giant magnon is identified with the momentum $p$ of the magnon, which is therefore periodic \cite{Hofman:2006xt}. The limit of small $p$ such that $p\,\sqrt{\lambda}$ is kept fixed corresponds to the pp-wave regime \cite{Berenstein:2002jq}, while the intermediate region with fixed $p\,\sqrt[4]{\lambda}$ is called   ``near-flat space'' regime \cite{Arutyunov:2004vx, Maldacena:2006rv}.
Similar classical string configurations are called single spike strings: they have a finite angular amplitude in one direction of the sphere as the giant magnons, but they wind infinitely many times around the orthogonal angular direction \cite{Ishizeki:2007we, Mosaffa:2007ty, Ishizeki:2007kh}. Their interpretation from the gauge theory side is still unclear.

Despite the deep progress achieved in understanding the integrable structure of the duality between type IIB string theory on $AdS_5 \times S^5$ and  $\mathcal{N}=4$ SYM, not many results are available about integrability in the four dimensional AdS/CFT dualities with less symmetries. The example mainly explored in this direction is the $\beta$ deformation of $AdS_5 \times S^5$, whose background has been found in \cite{Lunin:2005jy} and the dual gauge theory is the $\mathcal{N}=1$ SCFT obtained through the marginal deformation of the $\mathcal{N}=4$ \cite{Leigh:1995ep}. 
The background is obtained by applying to $AdS_5 \times S^5$ a TsT transformation (i.e. a T duality in one angle, a shift of another isometry variable and another T duality in the first angle) containing the deformation parameter. 
Giant magnons and spiky strings have been studied on this background \cite{Chu:2006ae, Bobev:2006fg, Bobev:2007bm, Bykov:2008bj}.
The dispersion relation for a single magnon turns out to be the one of the sphere with the momentum shifted by a quantity which contains the deformation parameter linearly\footnote{Another background whose dual gauge theory is $\mathcal{N}=1$ is the Maldacena-Nunez solution \cite{Maldacena:2000yy}, but the integrability aspects are less studied here. For the near-flat space limit on this background, see \cite{Kreuzer:2007az}.}.

Another very important class of AdS/CFT dualities with $\mathcal{N}=1$ dual gauge theories has been obtained by considering type IIB on $AdS_5 \times M^5$, where the internal $M^5$ is a compact Einstein-Sasaki manifold, for which the minimal supersymmetry is preserved. The simplest example of such manifolds is $M^5=T^{1,1}$, whose $\mathcal{N}=1$ dual gauge theory has been proposed in \cite{Klebanov:1998hh}. This geometry arises from the near-horizon limit of a stack of $N$ D3 branes placed at the conifold singularity. 
After this example, infinite families of five dimensional Einstein-Sasaki spaces \cite{Gauntlett:2004yd, Cvetic:2005ft} and their dual gauge theories \cite{Benvenuti:2004dy, Benvenuti:2005ja, Franco:2005sm, Butti:2005sw} have been constructed. The pp-wave limit for these geometries has been considered in \cite{Itzhaki:2002kh, Gomis:2002km, Pando Zayas:2002rx, Kuperstein:2006xw}, while their near-flat space limit has been studied in \cite{Benvenuti:2007qt}.\\
As for type IIB on $AdS_5 \times T^{1,1}$, computations have been done on this background, in order to understand the spectrum of the chiral primaries of the dual gauge theory \cite{Gubser:1998vd,Kim:2003vn}.

In this paper we find giant magnons and spiky strings solutions moving in a sector of $AdS_5 \times T^{1,1}$.
In particular, we consider classical spinning strings moving in the background $\mathbb{R}\times \Sigma_b$, where $t \in \mathbb{R}$ and $\Sigma_b$ is the three dimensional squashed sphere ($b$ is the squashing parameter). For $b=2/3$, it becomes the consistent sector of  $\mathbb{R}\times T^{1,1}$ given by constant $(\phi_2,\theta_2)$ (namely we shrink to a point one of the two $S^2$'s occurring in the metric of $T^{1,1}$); while for $b=1$ it reduces to $\mathbb{R} \times S^3$, the well known sector of $AdS_5 \times S^5$. In order to understand the differences between these two special cases,
we work with generic $b \in (0,1]$. \\
By writing explicitly the solutions for the giant magnons and the single spike strings on $\mathbb{R}\times \Sigma_b$, we can compute the finite combinations of their energy and momenta, finding also the dispersion relation of the giant magnons and the single spike strings on the squashed three dimensional sphere. Then, such dispertion relations can be specialized to the sector of $AdS_5 \times T^{1,1}$ we are interested in and to $\mathbb{R}\times S^3 \subset AdS_5 \times S^5$ as well.\\
For the giant magnons on the conifold, we get the following dispersion relation
\begin{equation}
\label{intro conifold}
\frac{\sqrt{3}}{2}(\mathcal{E}-3\,\mathcal{J}_\phi)\,=\,
\frac{\cos\big(\sqrt{3}\,(\mathcal{E}-3\,\mathcal{J}_\psi)/2\big)-\cos \Delta \varphi}{\sin\big(\sqrt{3}\,(\mathcal{E}-3\,\mathcal{J}_\psi)/2\big)}
\end{equation}
where $\Delta\varphi$ is the finite angular amplitude of the giant magnons. 
Instead, in the limit $b\rightarrow 1$, our dispersion relation reduces to
\begin{equation}
\label{intro S3}
(\mathcal{E}-\mathcal{J}_1)^2-
\mathcal{J}_2^2
\,=\,
4 \sin^2\frac{\Delta \varphi_1}{2}
\end{equation}
which is the known dispersion relation of the giant magnons on $S^3$ \cite{Dorey:2006dq}.
The qualitative difference between (\ref{intro conifold}) and (\ref{intro S3}) is that on the conifold also the energy and the momenta enter in the dispersion relation in a transcendental way, while on the sphere they occur just quadratically.\\
Our results could be helpful also to study the example of non-critical eight dimensional AdS/CFT considered in \cite{Bigazzi:2005md}, where the squashed three dimensional sphere with $b=2/3$ occurs in the compact part.
We notice that for $b=1/2$ our background gives a submanifold of $AdS_4 \times Q^{1,1,1}$ \cite{D'Auria:1983vy}, therefore our results could be useful also in the context of $AdS_4/CFT_3$.

The paper is organized as follows. In the section \ref{GM and SS}, after having introduced the ansatz for the solution on the squashed three dimensional sphere, we discuss the differential equations characterizing the problem and the boundary conditons corresponding to the giant magnon and single spike string solution. After having written the solutions, we find the corresponding angular amplitudes.
In the section \ref{section dispersion relation GM} we focus on the giant magnon solution, finding its finite combinations of global charges and the dispersion relation occurring among them. 
The limits of pointlike string (BMN, \cite{Berenstein:2002jq}) and folded string (GKP, \cite{Gubser:2002tv}) of this relation are also considered.
An important special case we study is given by the ``basic'' giant magnons, which correspond to the giant magnons on $\mathbb{R}\times S^2$, originally found in \cite{Hofman:2006xt}. This is done for a generic value of the squashing parameter $b$; therefore one can check for every step that the limit  $b \rightarrow 1$ provides the corresponding well known result on the three dimensional sphere. In the section \ref{section dispersion relation SS} we give the finite combinations of global charges and the dispersion relation for the single spike string while in the section \ref{section gauge theory} we discuss the gauge operator dual to the giant magnon.\\
In the appendix \ref{appendix sector T11} we show that the squashed three dimensional sphere for the special value of the squashing parameter $b=2/3$ is a consistent sector of $AdS_5 \times T^{1,1}$.
In the appendix \ref{appendix symmetries} we briefly discuss the symmetries of $T^{1,1}$. 
In the appendix \ref{appendix known equations} we show how the main differential equation we deal with in our problem is related to the double sine-Gordon equation and to the compound KdV equation.
In the appendix \ref{appendix 2 remarks} we discuss the solution of our differential equation in two special regimes, while in the appendix \ref{appendix integrals} we explicitly write the expressions of some useful integrals occurring in the computation of the conserved charges. In the appendix \ref{appendix spin chain} we perform the large spin limit introduced in \cite{Kruczenski:2003gt} for our background, in order to understand a possible relation between the sigma model and a spin chain.

\section{Giant magnons and single spiky strings on $\mathbb{R}\times \Sigma_{b}$}
\label{GM and SS}

In this section we consider classical strings on $\mathbb{R}\times \Sigma_{b}$, where $\Sigma_{b}$ is the squashed three dimensional sphere. First we introduce the ansatz for the solution and write down the key equation to solve. Then, after having discussed its boundary conditions, we explicitly find the solutions corresponding to the giant magnons and the single spiky strings on $\mathbb{R}\times \Sigma_{b}$.

In the appendix \ref{appendix sector T11} we have shown that the submanifold given by either $(\theta_1,\phi_1)=\textrm{const}$ or $(\theta_2,\phi_2)=\textrm{const}$ is a consistent sector of classical string theory on $\mathbb{R}\times T^{1,1}$. 
Here we consider a more general situation by taking the Polyakov action and the Virasoro constraints on  $\mathbb{R}\times \Sigma_{b}$, namely with the following target space metric
\begin{equation}
\label{metric b}
ds^2\,=\,-\,dt^2+\,\frac{b}{4}\,\Big( b\,\big(d\psi\,-\,\cos\theta\,d\phi \big)^2+\,d\theta^2+\,\sin^2\theta\,d\phi^2 \Big)
\end{equation}
where $t\in \mathbb{R}$ is the time coordinate of $AdS_5$, while $\psi\in[0,4\pi)$, $\theta\in[0,\pi]$ and $\phi\in [0,2\pi)$ are the angular coordinates of the squashed three dimensional sphere.\\
The advantage of using (\ref{metric b}) is that for $b=2/3$ we get the consistent sector of $\mathbb{R} \times T^{1,1}$ described in the appendix \ref{appendix sector T11}, while for $b=1$ the metric (\ref{metric b}) becomes the well known metric of $\mathbb{R} \times S^{3}$. 
In order to compare with the previous results, in the $b=1$ case it is better to perform the following change of coordinates
\begin{equation}
\label{S3coord}
\theta\,=\,2\,\eta
\hspace{1.5cm}
\psi\,=\,\varphi_1+\,\varphi_2
\hspace{1.5cm}
\phi\,=\,\varphi_1-\,\varphi_2
\end{equation} 
where the ranges for these angular variables are $\eta \in [0,\pi/2]$, $\varphi_1\in [0,2\pi)$ and $\varphi_2\in [0,2\pi)$.\\
The target space metric (\ref{metric b}) has three isometries corresponding to the shifts of $t$, $\psi$ and $\phi$ by independent constants. The momenta associated to these isometries read
\begin{eqnarray}
\label{pt}
& & p_t    \;=\;-\,T  \, \partial_\tau t
\\
\label{ppsi}
\rule{0pt}{.8cm}
& & p_\psi \;=\; T \, \frac{b^2}{4} \big( \partial_\tau \psi -\cos\theta\,\partial_\tau \phi \big) \\
\label{pphi}
\rule{0pt}{.8cm}
& & p_\phi  \;=\;T\, \frac{b}{4}\,\Big(\big(\,b\,\cos^2\theta+\sin^2\theta\,\big) \partial_\tau \phi - b\,\cos\theta\,\partial_\tau \psi\Big) \phantom{xxxx}
\end{eqnarray}
and the corresponding conserved charges are
\begin{equation}
\label{charges}
E\,=\,-\int d\sigma\, p_t
\hspace{1.5cm}
J_\psi\,=\,\int d\sigma\, p_\psi
\hspace{1.5cm}
J_\phi\,=\,\int d\sigma\, p_\phi\;.
\end{equation} 
In the following we mainly use $\mathcal{E} \equiv E/T$, $\mathcal{J}_\psi \equiv J_\psi/T$ and $\mathcal{J}_\phi \equiv J_\phi/T$.\\
In order to find classical solutions for the sigma model characterized by (\ref{metric b}), one introduces the following ansatz
\begin{equation}
\label{GMansatz}
t\,=\,k\,\tau 
\hspace{1.5cm}
\theta\,=\,\theta(y)
\hspace{1.5cm}
\psi\,=\,\omega_\psi\,\tau\,+\,\Psi(y)
\hspace{1.5cm}
\phi\,=\,\omega_\phi\,\tau\,+\,\Phi(y)
\end{equation} 
where $y=c\,\sigma-d\,\tau$, $\omega_\psi$ and $\omega_\phi$ are constants, and  $\theta(y)$, $\Psi(y)$ and $\Phi(y)$ are the functions that we have to find. Spinning strings on $\mathbb{R} \times T^{1,1}$ have been considered also in \cite{Kim:2003vn}, but closed strings configurations have been analyzed there and a simpler ansatz for the solutions has been used.\\
Given the ansatz (\ref{GMansatz}), the equation of motion for $t$ is trivially satisfied. The equations of motion coming from the variation of the action w.r.t.  $\psi$ and $\phi$, once written in terms of $y$, can be integrated once, providing respectively $\Psi'(y)$ and $\Phi'(y)$ in terms of $\theta(y)$ as follows
\begin{eqnarray}
\label{Psiprime}
\Psi'& = &
\frac{1}{c^2-d^2}
\left[\,\frac{4}{b\,\sin^2\theta}\left(A_\psi\left(\cos^2\theta+\frac{\sin^2\theta}{b}\,\right)+A_\phi \cos\theta \right)
 - d\,\omega_\psi\,\right]
\phantom{xxxi}
\\
\label{Phiprime}
\rule{0pt}{.9cm}\Phi'& = &
\frac{1}{c^2-d^2}
\left[\,\frac{4}{b\,\sin^2\theta}\big( A_\psi \cos\theta+A_\phi \big)
 - d\,\omega_\phi\,\right]
\end{eqnarray}
where $A_\psi$ and $A_\phi$ are the integration constants. Then, by plugging the above expressions for $\Psi'$ and $\Phi'$ into the Virasoro constraint (\ref{VC2}) written in terms of $y$, we get the following equation for $\theta(y)$ 
\begin{eqnarray}
\label{thetaEq}
(\theta')^2& = & \frac{4}{b\,(c^2-d^2)^2}
\left[\,
\frac{c^2+d^2}{d}\,\big(A_\psi\,\omega_\psi+A_\phi\,\omega_\phi \big)
\right.\\
\rule{0pt}{.6cm}
& & \hspace{2.5cm}
-\;c^2\,\frac{b}{4}\,\Big(b\,\omega_\psi^2-2\,b\,\omega_\psi \,\omega_\phi \,\cos\theta
+\big(\,b\, \cos^2\theta+\sin^2\theta\,\big)\,\omega_\phi^2 \Big)
\nonumber\\
\rule{0pt}{.75cm}
& & \hspace{2.5cm}
\left.
-\;\frac{4}{b\,\sin^2\theta}\left(A_\psi^2\left(\cos^2\theta+\frac{\sin^2\theta}{b}\,\right)
+2\,A_\psi\,A_\phi \cos\theta +A_\phi^2\,\right)\,
\right]. \nonumber
\end{eqnarray}
It is crucial to observe that the differential equation for $\theta(y)$ coming from the equation of motion for $\theta$ is already encoded in (\ref{thetaEq}), since it can be recovered by deriving (\ref{thetaEq}) w.r.t. $y$ and assuming $\theta' \neq 0$.
Finally, by employing (\ref{Psiprime}), (\ref{Phiprime}) and (\ref{thetaEq}), the Virasoro constraint (\ref{VC1}) reduces to the following algebraic relation
\begin{equation}
\label{k^2}
k^2\,=\,\frac{\omega_\psi\,A_\psi +\omega_\phi\,A_\phi }{d}\;.
\end{equation}
Thus, the differential equation we have to solve is (\ref{thetaEq}). Given its solution $\theta(y)$, one can find $\Psi(y)$ and  $\Phi(y)$ by integrating once (\ref{Psiprime}) and (\ref{Phiprime}) respectively.\\
Introducing $u \equiv \cos^2(\theta/2)$,
the equation (\ref{thetaEq}) becomes
\begin{equation}
\label{total energy full}
\frac{(u')^2}{2}\,+\,V(u)
\,=\,0
\hspace{.7cm}
\textrm{with}
\hspace{.7cm}
V(u)\,=\,-\,2 \left(\alpha_8\, u^4 + \alpha_6\, u^3 + \alpha_4\, u^2+ \alpha_2\, u\,+ \alpha_0 \right)
\end{equation}
where the coefficients read
\begin{eqnarray}
& &
\hspace{-.7cm}
\alpha_8 \,=\,   -\,\frac{c^2 (1-b)\,\omega_\phi^2}{(c^2-d^2)^2} \\
\rule{0pt}{.8cm}
& &
\label{alpha 6}
\hspace{-.7cm}
\alpha_6 \,=\,  \frac{c^2 \big(2 (1-b)\,\omega_\phi - b\,\omega_\psi\big)\, \omega_\phi}{(c^2-d^2)^2}
\\
\rule{0pt}{.8cm}
& &
\label{alpha 4}
\hspace{-.7cm}
\alpha_4 \,=\, \frac{1}{4\,b\,(c^2-d^2)^2}  \left[\,\frac{16\,(1-b)\,A_\psi^2}{b^2}
- \frac{4\,(c^2+d^2)}{d}\,\big(A_\psi\,\omega_\psi+A_\phi\,\omega_\phi \big) \right.\\
& &\hspace{7cm}
\left.\phantom{\frac{1}{2}}
+b\,c^2\big( \, b\,\omega_\psi^2 +6\,b\,\omega_\phi\,\omega_\psi+(5\,b-4)\,\omega_\phi^2\,\big)
\right]
\nonumber \\
\rule{0pt}{.7cm}
& &
\hspace{-.7cm}
\alpha_2 \,=\,  \frac{1}{4\,b\,(c^2-d^2)^2}  \left[\, - \,\frac{16\,A_\psi \,((1-b)\,A_\psi+b\,A_\phi )}{b^2}
+ \frac{4\,(c^2+d^2)}{d}\,\big(A_\psi\,\omega_\psi+A_\phi\,\omega_\phi \big)\right.\\
& &\hspace{11cm}
\left. \phantom{\frac{1}{2}}
- b^2  c^2 \big(\omega_\psi+\omega_\phi\big)^2\right] \nonumber
\\
\rule{0pt}{.8cm}
& &
\hspace{-.7cm}
\alpha_0 \,=\,   -\,\frac{(A_\psi - A_\phi)^2}{b^2(c^2-d^2)^2}\;.
\end{eqnarray}
The equation (\ref{total energy full}) can be interpreted as the conservation of energy
for the motion governed by the quartic potential $V(u)$. 
Notice that in our case $\alpha_8 \leqslant 0$ (being $0<b\leqslant 1$) and $u(y) \in [0,1]$.
We remark also that in the limit $b \rightarrow 1$, where the target space reduces to $\mathbb{R}\times S^3$, we have that $\alpha_8 \rightarrow 0$ and therefore the potential becomes cubic.

\subsection{Boundary conditions}

In order to find a solution to (\ref{total energy full}), we have to impose the boundary conditions.\\
Here we require that the open string configurations reach $\theta=\pi$ and that $\theta=\pi$ is a turning point, namely that $\theta'=0$ at $\theta= \pi$.
\\ 
From (\ref{Psiprime}), (\ref{Phiprime}) and (\ref{thetaEq}), we observe that the requirement of finiteness  for $\Psi'$, $\Phi'$ and $\theta'$ at $\theta=\pi$ leads to 
\begin{equation}
\label{Apsi=Aphi}
A_\psi \,=\,A_\phi\;.
\end{equation}
As for the condition that $\theta= \pi$ is a turning point, we expand the r.h.s. of (\ref{thetaEq}) around $\theta= \pi$ and, assuming (\ref{Apsi=Aphi}), we find that this condition provides the following equation
\begin{equation}
\label{eqA}
\frac{16}{b^2}\,A_\phi^2 
- \frac{4\,(c^2+d^2)}{d}\,\big(\omega_\psi+ \omega_\phi \big)\,A_\phi
+ b^2  c^2 \big(\omega_\psi+\omega_\phi\big)^2
\, = \, 0\;.
\end{equation}
Thus, our boundary conditions are given by (\ref{Apsi=Aphi}) and (\ref{eqA}).
After having imposed them, the r.h.s. of (\ref{thetaEq}) is $O((\theta-\pi)^2)$ 
and we have $\alpha_0 = 0$ and $\alpha_2 = 0$ in (\ref{total energy full}).
The solutions of (\ref{eqA}) are
\begin{eqnarray}
\label{AGM}
& &
A_\phi\,=\,\frac{b^2}{4}\,d\,(\omega_\psi+\omega_\phi )
\hspace{2.5cm}
\textrm{giant magnons}
\\
\label{ASS}
\rule{0pt}{.9cm}
& &
A_\phi\,=\,\frac{b^2}{4}\;\frac{c^2\,(\omega_\psi+\omega_\phi )}{d}
\hspace{2.3cm}
\textrm{single spike strings}
\end{eqnarray}
and they characterize two distinct classical string configurations: the giant magnons and the single spike strings respectively.
\noindent  From (\ref{k^2}), we find the corresponding values for $k^2$
\begin{eqnarray}
\label{k2GM}
& & 
k^2\,=\,b^2 \left(\frac{\omega_\psi+\omega_\phi}{2}\right)^2
\hspace{2.2cm}
\textrm{giant magnons}
\\
\label{k2SS}
\rule{0pt}{.9cm}
& &
k^2\,=\,b^2\,\frac{c^2}{d^2} \left(\frac{\omega_\psi+\omega_\phi}{2}\right)^2
\hspace{1.75cm}
\textrm{single spike strings}\,.
\end{eqnarray}
Since the boundary conditions (\ref{Apsi=Aphi}) and (\ref{eqA}) imply $\alpha_0=\alpha_2 =0$ and the expressions for $\alpha_8$ and $\alpha_6$ do not contain the integration constants $A_\psi$ and $A_\phi$, only $\alpha_4$ distinguishes between the giant magnon and the spiky string solutions. 
Plugging (\ref{Apsi=Aphi}) and respectively (\ref{AGM}) and (\ref{ASS}) into (\ref{alpha 4}), one finds the expressions of $\alpha_4$ for the giant magnons and the single spiky strings.
Since $c^2-d^2$ is positive for the giant magnons and negative for the single spike strings, it is convenient to introduce $v^2 \equiv d^2/c^2$ for the giant magnons and $v^2 \equiv c^2/d^2$ for the single spike strings. Then, the two expressions for $\alpha_4$ assume the same form
\begin{equation}
\label{alpha4 common}
\alpha_4\,=\,
\frac{c^2 \omega_\phi^2}{4\,(c^2-d^2)^2} \,\hat{\alpha}_4
\hspace{1cm}
\textrm{where}
\hspace{1cm}
\hat{\alpha}_4\,\equiv\,
4 ( b\,\Omega - 1+b)
- v^2 b^2 (\Omega+1)^2
\end{equation}
where $\Omega \equiv \omega_\psi /\omega_\phi$. 
Thus, the boundary conditions (\ref{Apsi=Aphi}) and (\ref{eqA}) remarkably simplify the differential equations we are dealing with. In particular, (\ref{total energy full}) reduces to 
\begin{equation}
\label{total energy GM}
\frac{(u')^2}{2}\,+\,V(u)
\,=\,0
\hspace{.7cm}
\textrm{with}
\hspace{.7cm}
V(u)\,=\,-\,2\,u^2 \left(\alpha_8\, u^2 + \alpha_6\, u + \alpha_4 \right)
\end{equation}
while for (\ref{Psiprime}) and (\ref{Phiprime}) we get
\begin{eqnarray}
\label{Psiprime 1}
& & \hspace{-.8cm}
\Psi' \,=\,
\frac{1}{b^2(c^2-d^2)}
\left(\,\frac{2(2-b)A_\phi-d\,b^2\omega_\psi}{1-u}
-\frac{4(1-b)A_\phi-d\,b^2\omega_\psi}{1-u}\,u
\right) \\
\rule{0pt}{.8cm}
\label{Phiprime 1}
& & \hspace{-.8cm}
\Phi' \,=\,
\frac{1}{b\,(c^2-d^2)}
\left(\,\frac{2\,A_\phi-d\,b\,\omega_\phi}{1-u}
+\frac{d\,b\,\omega_\phi}{1-u}\,u
\right)\;.
\end{eqnarray}
The first differential equation we have to solve is (\ref{total energy GM}), where
we remind that $\alpha_8 \leqslant 0$ and $u(y) \in [0,1]$. The potential has a double zero in $u=0$, where $\partial_u^2 V(0) = -\,4\,\alpha_4$, while the other two zeros are the roots $\alpha_< \leqslant \alpha_>$ of the second order polynomial $\alpha_8\, u^2 + \alpha_6\, u + \alpha_4$.\\

\noindent {\bf The condition $\alpha_4>0$.} Now we require $\alpha_4>0$ and we will consider this case for the remaining part of the paper. The solution for $\alpha_4<0$ and $\alpha_4=0$ is briefly discussed in the appendix \ref{appendix 2 remarks}, but it is meaningless for us, since it does not satisfy our boundary conditions.\\
Since $\alpha_8 \leqslant 0$, the condition $\alpha_4 > 0$ implies that the roots $\alpha_<$ and $\alpha_>$ are real and also that $\alpha_<  <  0  < \alpha_>$, where
\begin{equation}
\label{positive alpha}
\alpha_>\,=\,-\,\frac{\alpha_6 + \sqrt{\alpha_6^2-4\,\alpha_4\,\alpha_8}}{2\,\alpha_8}
\,=\,
\frac{2(1-b)-b\,\Omega+b\,\sqrt{\Omega^2-(1-b)\,v^2(1+\Omega)^2}}{2(1-b)}
\;.
\end{equation}
The differential equation in (\ref{total energy full}) with the potential (\ref{total energy GM}) can be written as
\begin{equation}
\label{total energy GM bis}
(u')^2\,=\,\frac{4\,c^2(1-b)\omega_\phi^2}{(c^2-d^2)^2}\;u^2 (\alpha_> -u)(u-\alpha_<)
\end{equation}
therefore when $\alpha_4>0$ the motion is limited to the range $u \in [0,\alpha_>]$, where $u=0$ and $u=\alpha_>$ are the turning points. Since $u=\cos^2(\theta/2) \leqslant 1$, a physical condition for the solution is $\alpha_> \leqslant 1$.
\begin{figure}[t]
\begin{tabular}{ccc}
\includegraphics[width=7.4cm]{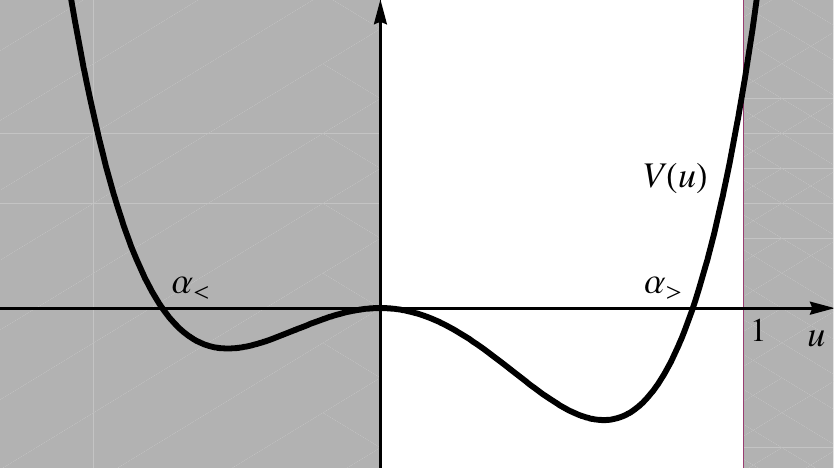}
& \hspace{.1cm} & 
\includegraphics[width=7.4cm]{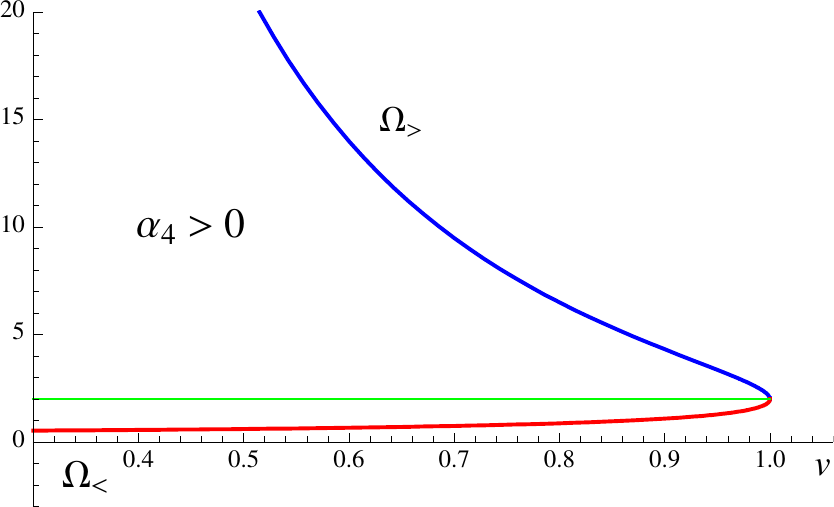}
\end{tabular}
\caption{on the left, the potential $V(u)$ in (\ref{total energy GM}) with $\alpha_4 > 0$. On the right, the region $\alpha_4>0$ given by the bounds  (\ref{bounds Omega}) in the case of the conifold. The horizontal green line corresponds to the ratio (\ref{ratio 0}).
\label{picture potential GM}}
\end{figure}
From (\ref{positive alpha}), one finds that $\alpha_> \leqslant 1$ is equivalent\footnote{Notice that $-\,2\alpha_8-\alpha_6>0$ when $\omega_\psi\, \omega_\phi >0$.} to $\alpha_8 + \alpha_6 + \alpha_4 \leqslant 0$.
Substituting the explicit expressions for $\alpha_8$, $\alpha_6$ and $\alpha_4$, and using the boundary conditions (\ref{Apsi=Aphi}) and (\ref{eqA}), we find
\begin{equation}
\label{alpha8 - alpha6 -alpha4}
\alpha_8 + \alpha_6 + \alpha_4 \,=\,
-\,\frac{4\,A_\phi^2}{b^2(c^2-d^2)^2}
\end{equation}
where $A_\phi$ is given in (\ref{AGM}) and (\ref{ASS}) for the cases we are considering. 
Thus, the condition $\alpha_> \leqslant 1$ is always satisfied. 
Notice that for $v\rightarrow 0$, we have $\alpha_>  \rightarrow 1$ and the string extends between $\theta=0$ and $\theta=\pi$. The potential we are dealing with is shown in the picture on the left of the figure \ref{picture potential GM}, where the grey regions are physically meaningless.\\
\noindent The condition $\alpha_4 >  0$ constrains the choice of $\Omega = \omega_\psi/\omega_\phi$. Assuming $\omega_\phi > 0$, it gives the following bounds
\begin{equation}
\label{bounds Omega}
\Omega_<\,\equiv\,
\frac{2\,-b\,v^2-2\, \sqrt{1-v^2}}{b\,v^2}\,
\,<\,\Omega\,<\,
\frac{2\,-b\,v^2+2\, \sqrt{1-v^2}}{b\,v^2}
\,\equiv\,\Omega_>
\end{equation}
where $v \in (0,1]$. Notice that the ratio 
\begin{equation}
\label{ratio 0}
\Omega
\,=\,
\frac{2-b}{b}
\end{equation}
is the only constant value which is allowed for any $v \in [0,1]$. 
When $v \rightarrow 0$ the bound (\ref{bounds Omega}) becomes $(1-b)/b < \Omega < +\infty$.
In the picture on the right of the figure \ref{picture potential GM}, the region $\alpha_4 >  0$ and the ratio (\ref{ratio 0}) are shown in the case of the conifold (i.e. for $b=2/3$). 

\subsection{Explicit solution}
\label{subsection explicit solution}

Now we discuss the solution of (\ref{total energy GM}) with $\alpha_4>0$.
We are looking for  $u(y)$ defined for $y\in \mathbb{R}$ and such that $u$, $u'$, $u'' \rightarrow\,0$ as $y \rightarrow \pm \infty$ (solitary wave solution).
It is given by\footnote{The argument of cosh can be shifted by an arbitrary constant, but here we choose the solution invariant under $y\rightarrow -\,y$, for which this constant is zero.}
\begin{equation}
\label{solution u}
u(y)\,=\,\frac{\gamma}{\cosh ( 2 \sqrt{\alpha_4}\,y )-R}
\end{equation}
where
\begin{eqnarray}
\label{gamma}
\gamma & \equiv &
\frac{2\,\alpha_4}{\sqrt{\alpha_6^2-4\,\alpha_4\,\alpha_8}}
\;=\;
\frac{4(b\,\Omega - 1+b)
- v^2 b^2 (\Omega+1)^2}{2\,b\, \sqrt{\Omega^2-(1-b)\,v^2(1+\Omega)^2}}
\\
\rule{0pt}{.7cm}
\label{R}
R &\equiv&
\frac{\alpha_6}{\sqrt{\alpha_6^2-4\,\alpha_4\,\alpha_8}}
\;=\;
\frac{2(1-b)-b\,\Omega}{b\, \sqrt{\Omega^2-(1-b)\,v^2(1+\Omega)^2}}\;.
\end{eqnarray}
We remark that $\alpha_4 > 0$ guarantees both $\gamma>0$ and $|R| < 1$, therefore (\ref{solution u}) is consistent with $u \geqslant 0$ and it is also a well defined solitary wave solution.
Moreover, $\alpha_4 > 0$ and $\alpha_8 + \alpha_6 + \alpha_4 < 0$ imply $|R+\gamma| < 1$. We recall that only $\alpha_4$ distinguishes between the giant magnons and the single spike strings.\\
At this point, we find it convenient to introduce the following parameterization
\begin{eqnarray}
\label{beta def}
R+\gamma & \equiv & -\,\cos \beta  \hspace{2cm} \beta  \in [0,\pi]\\
\label{delta def}
R & \equiv & -\,\cos \delta \hspace{2.1cm} \delta  \in [0,\pi]\ \;.
\end{eqnarray}
Notice that $\gamma \geq 0$ implies $\delta \leq \beta$. Since $\hat{\alpha}_4$, $\gamma$ and $R$, given in (\ref{alpha4 common}), (\ref{gamma}) and (\ref{R}) respectively, depend on $\Omega$ and $v$ only, a relation must occur among them. It reads
\begin{equation}
\frac{\hat{\alpha}_4}{4(1-b)}
\,=\,
\frac{\gamma^2}{1-R^2}
\end{equation}
which can be written also in term of $\beta$ and $\delta$ as
\begin{equation}
\label{alpha4 relation 1}
\frac{\sqrt{\hat{\alpha}_4}}{2 \sqrt{1-b}} \,=\,
 \frac{\cos\delta-\cos\beta}{\sin{\delta}}\;.
\end{equation}
This relation will be useful to find the dispersion relation of the giant magnons and the single spike strings (see the sections \ref{section dispersion relation GM} and \ref{section dispersion relation SS}).

Now we find it convenient to consider also the regime of small $\beta$, $\delta$ and $\hat{\alpha}_4$. First, one observes that, assuming $\alpha_6<0$, we have $\delta^2=O(\hat{\alpha}_4)$ and $\beta^2=O(\hat{\alpha}_4)$. Then, the expansion of (\ref{alpha4 relation 1}) to the first non trivial order gives the following relation
\begin{equation}
\label{alpha4 small}
\delta\left(\delta+\frac{\sqrt{\hat{\alpha}_4}}{\sqrt{1-b}} \right) \,=\,
\beta^2
\end{equation}
which will be useful for BMN limit of the solution (see the subsection \ref{BMN and GKP}).\\

\noindent {\bf The special case of the three sphere.} Taking the limit $b \rightarrow 1$ of our solution, we can recover the well known results on  $\mathbb{R}\times S^3$. \\
Denoting by $(\eta, \varphi_1, \varphi_2)$ the angular coordinates on $S^3$, from the ansatz (\ref{GMansatz}) and the change of coordinates (\ref{S3coord}), we have $\eta = \eta(y)$, $\varphi_1\,=\,\omega_1\,\tau\,+\,\Phi_1(y)$ and $\varphi_2\,=\,\omega_2\,\tau\,+\,\Phi_2(y)$,
where $\omega_1=(\omega_\psi+\omega_\phi)/2$, $\omega_2=(\omega_\psi-\omega_\phi)/2$ and $\Phi_1=(\Psi+\Phi)/2$, $\Phi_2=(\Psi-\Phi)/2$.\\
As for the explicit solution for $\eta(y)$, taking $\omega_1^2 > \omega_2^2$ and $\alpha_4|_{b=1} > 0$, from (\ref{gamma}) and (\ref{R}) one can see that for $b=1$ we have $\gamma= 2\,\alpha_4/|\alpha_6|>0$ and $R=-1$. Thus, for $b=1$ the solution (\ref{solution u}) reduces to
\begin{equation}
\label{solutionS3}
\cos \eta(y)\,=\,
\frac{\sqrt{\alpha_4/|\alpha_6|}}{\cosh\big(\sqrt{\alpha_4}\, y\big)}
\end{equation}
where $\alpha_4$ and $\alpha_6$ are evaluated for $b=1$. Notice that in this limit we have $\delta|_{b=1}=0$. The other relations for the giant magnons and the single spike strings on $\mathbb{R}\times S^3$ coming from the remaining equations of motion and the Virasoro constraints are recovered in this limit as well.

\subsection{\bf Angular amplitudes of the solution}

Given the solution (\ref{solution u}) with $\alpha_4 >0$, it is straightforward to check that $\max_{y\,\in\, \mathbb{R}} (u)\equiv \cos^2(\tilde{\theta}/2)= \gamma/(1-R)=\alpha_> $, which corresponds to the minimum value $\tilde{\theta}$ assumed by $\theta$.
In other words, the solution we are considering has $\tilde{\theta} \leqslant \theta \leqslant \pi$ and therefore their angular amplitude $\Delta\theta \equiv \pi-\tilde{\theta}$ along the direction of $\theta$ is
\begin{equation}
\label{Delta theta b}
\Delta\theta \,=\,
\pi-2\,\arctan\sqrt{\frac{1-R-\gamma}{\gamma}}\;.
\end{equation}
To compute the angular amplitudes of the solutions along the directions of $\psi$ and $\phi$, we have to integrate (\ref{Psiprime 1}) and (\ref{Phiprime 1}) over $y \in \mathbb{R}$.
Since $u=u(y)$ is given by (\ref{solution u}), the integral of $1/(1-u)$ over $y \in \mathbb{R}$ is divergent while the one of $u/(1-u)$ converges; therefore, since the coefficients of $1/(1-u)$ in (\ref{Psiprime 1}) and (\ref{Phiprime 1}) do not vanish neither for (\ref{AGM}) nor for (\ref{ASS}), both $\Delta\psi$ and $\Delta\phi$ are divergent quantities for general allowed values of $\omega_\psi$ and $\omega_\phi$. Instead, it is useful to introduce the following combination
\begin{equation}
\label{DeltaPhib}
\Delta \varphi \,\equiv\,\frac{\Delta \psi + \Delta \phi}{2}\,=\,
\int_{-\infty}^{+\infty}\,
\frac{\Psi' + \Phi'}{2}\,dy\;.
\end{equation}
This angular amplitude is finite for the giant magnons, but it remains divergent for the single spike strings. Notice that for $b=1$, we have that $\Delta\varphi|_{b=1}=\Delta\varphi_1$ becomes the angular amplitude of the solution in the direction of the coordinates $\varphi_1$ on $S^3$.

\section{Giant magnons: energy and momenta}
\label{section dispersion relation GM}

In this section we compute the conserved charges for the giant magnon solution and we write the corresponding dispersion relation for a generic value of the squashing parameter. The limits of pointlike string (BMN regime \cite{Berenstein:2002jq}) and folded string (GKP regime \cite{Gubser:2002tv}) are also considered.

The giant magnon solution is characterized by the value (\ref{AGM}) for $A_\phi = A_\psi$. Choosing the positive root for $k$  in (\ref{k2GM}), the three momenta (\ref{pt}), (\ref{ppsi}) and (\ref{pphi}) for the giant magnon solution become
\begin{eqnarray}
\label{ptGM}
p_t &=& -\,T  \; \frac{b}{2}\,(\omega_\psi+\omega_\phi) 
\\
\label{ppsiGM}
\rule{0pt}{.8cm}
\frac{p_\psi}{b} 
&=&  T \left(\,\frac{b}{4}\,(\omega_\psi+\omega_\phi)-\frac{b\,\omega_\phi}{2(1-v^2)}\,u(y)\right) \\
\label{pphiGM}
\rule{0pt}{.8cm}
\frac{p_\phi}{b} 
&=&  T 
\left(\,\frac{b}{4}\,(\omega_\psi+\omega_\phi)
-\frac{\omega_\phi}{2(1-v^2)}\,\Big(2(1-b)\,u(y)^2
+\big(b\,\Omega-2(1-b)\big)\,u(y)\Big)\right)  
\phantom{xxx}
\end{eqnarray}
where $u(y)$ is given by (\ref{solution u}) with $v=d/c$. 
All the three charges $\mathcal{E}$, $\mathcal{J}_\psi$ and $\mathcal{J}_\phi$ diverge because 
of the constant term occurring in the corresponding momenta. Instead, the following linear combinations 
\begin{equation}
\mathcal{E}\,-\,\frac{2}{b}\,\mathcal{J}_\psi
\hspace{1.5cm}
\mathcal{E}\,-\,\frac{2}{b}\,\mathcal{J}_\phi
\hspace{1.5cm}
\mathcal{E}\,-\,\frac{\mathcal{J}_\psi+\mathcal{J}_\phi}{b}
\hspace{2cm}
\frac{\mathcal{J}_\psi-\mathcal{J}_\phi}{b}
\end{equation}
are finite, but only two of them are independent. By employing the integrals $U_1$ and $U_2$, given respectively in (\ref{U1}) and (\ref{U2}) of the appendix \ref{appendix integrals}, and the relation (\ref{U2 related to U1}) between them, we find that
\begin{equation}
\label{GM charges}
\mathcal{E}\,-\,\frac{2}{b}\,\mathcal{J}_\psi
\,=\,
\frac{b\,\delta}{\sqrt{1-b}}
\hspace{3cm}
\mathcal{E}\,-\,\frac{2}{b}\,\mathcal{J}_\phi
\,=\,
 \sqrt{\hat{\alpha}_4}
\end{equation}
which can be written in terms of $\Omega$ and $v$ only. We remark that $\mathcal{E}\,-2\,\mathcal{J}_\psi/b$ depends on $\Omega$ and $v$ in a transcendental way.\\
As for the angular amplitudes of the giant magnon solution, we have that $\Delta\psi$ and $\Delta\phi$ are separately divergent, but $\Delta\varphi$ is finite for any value of $\Omega$ in the allowed region (\ref{bounds Omega}).
Assuming $d(\omega_\psi+\omega_\phi)>0$, for $\Delta\varphi$ we find\footnote{In this computation the integral $\widetilde{U}$ given in (\ref{Utilde}) occurs.}
\begin{equation}
\label{DeltaPhib GM}
\Delta \varphi
 \,=\,
\beta\;.
\end{equation}
Thus, it turns out that the macroscopic quantities $\mathcal{E}-2\mathcal{J}_\psi/b$, $\mathcal{E}-2\mathcal{J}_\phi/b$ and $\Delta\varphi$ are related to the microscopic quantities $\delta$, $\hat{\alpha}_4$ and $\beta$ respectively, which can be expressed in terms of $\Omega$ and $v$ only. The relation (\ref{alpha4 relation 1}) occuring among the microscopic quantities can now be written in terms of the conserved charges as follows
\begin{equation}
\label{dispersion relation GM}
\frac{\mathcal{E}-2\,\mathcal{J}_\phi/b}{2\,\sqrt{1-b}}
\,=\,
\frac{\cos\big(\sqrt{1-b}\,(\mathcal{E}-2\,\mathcal{J}_\psi/b)/b\big)-\cos \Delta \varphi}{\sin\big(\sqrt{1-b}\,(\mathcal{E}-2\,\mathcal{J}_\psi/b)/b\big)}
\end{equation}
providing the dispersion relation of the giant magnons. Notice that, for a generic value of the squashing parameter, the finite combinations $\mathcal{E}-2\,\mathcal{J}_\psi/b$ and $\mathcal{E}-2\,\mathcal{J}_\phi/b$ enter in the dispersion relation in a transcendental way.\\
On the conifold (i.e. for $b=2/3$) the dispersion relation (\ref{dispersion relation GM}) becomes\footnote{In (\ref{dispersion relation conifold}) it is assumed that the conserved charges are evaluated for $b=2/3$ , while in (\ref{dispersion relation S3 from GM}) they are evaluated for $b=1$ as well.}
\begin{equation}
\label{dispersion relation conifold}
\frac{\sqrt{3}}{2}(\mathcal{E}-3\,\mathcal{J}_\phi)\,=\,
\frac{\cos\big(\sqrt{3}\,(\mathcal{E}-3\,\mathcal{J}_\psi)/2\big)-\cos \Delta \varphi}{\sin\big(\sqrt{3}\,(\mathcal{E}-3\,\mathcal{J}_\psi)/2\big)}
\end{equation}
and the transcendental dependence on $\mathcal{E}-3\,\mathcal{J}_\psi$ and $\mathcal{E}-3\,\mathcal{J}_\phi$ is still there.\\
Instead, by taking the limit $b\rightarrow 1$ of (\ref{dispersion relation GM}), we get
\begin{equation}
\label{dispersion relation S3 from GM}
(\mathcal{E}-2\,\mathcal{J}_\psi)\,
(\mathcal{E}-2\,\mathcal{J}_\phi)
\,=\,4 \sin^2\frac{\Delta \varphi_1}{2}
\end{equation}
which is the well known dispersion relation of the giant magnons on $\mathbb{R}\times S^3$ \cite{Dorey:2006dq}
\begin{equation}
\label{dispersion relation S3}
(\mathcal{E}-\mathcal{J}_1)^2-
\mathcal{J}_2^2
\,=\,
4 \sin^2\frac{\Delta \varphi_1}{2}
\end{equation}
once we use that $\mathcal{J}_1=(\mathcal{J}_\psi+\mathcal{J}_\phi)|_{b=1}$ and $\mathcal{J}_2=(\mathcal{J}_\psi-\mathcal{J}_\phi)|_{b=1}$. 
Notice that in this limit the dependence on the finite combinations of charges is not transcendental anymore, but just quadratic.
\begin{figure}[h]
\begin{center}
\includegraphics[width=7.8cm]{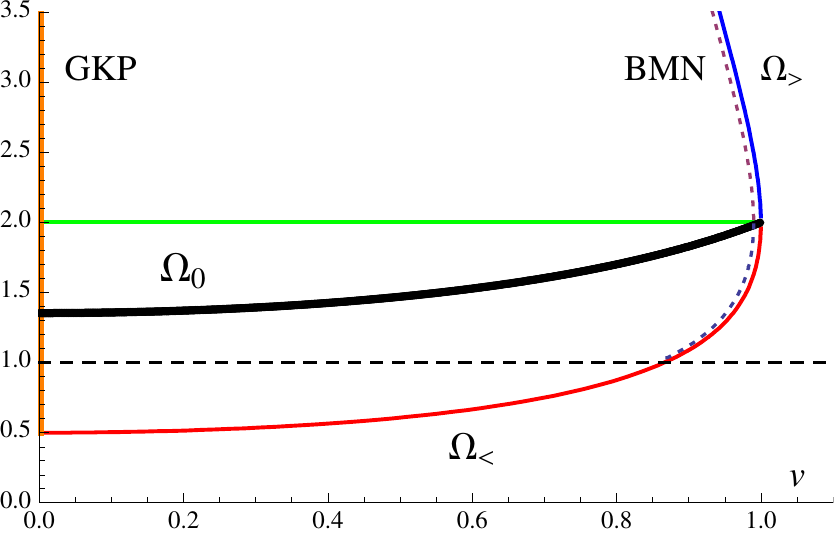}
\end{center}
\caption{The region $\alpha_4>0$ for the conifold (see the picture on the right of the figure \ref{picture potential GM}) with some interesting special cases. The black line shows the function $\Omega_0(v)$ for the ``basic'' giant magnons (i.e. the solution of (\ref{trans eq})), for which $\mathcal{J}_\psi=\mathcal{J}_\phi$. The horizontal dashed line corresponds to $\alpha_6=0$ and $\alpha_6 <0$ above it. The $v = 0$ region corresponds to the folded string limit (GKP), while the dotted line shows the region of the pointlike string limit (BMN).
\label{zoom bound omega picture}}
\end{figure}

\subsection{The BMN and GKP regimes}
\label{BMN and GKP}

Now we consider the dispersion relation (\ref{dispersion relation GM}) in the two limits of pointlike string (BMN, \cite{Berenstein:2002jq}) and folded string (GKP, \cite{Gubser:2002tv}).\\
{\bf The BMN limit.} In the BMN regime we have $\Delta \phi \sim \mathcal{E}-2\,\mathcal{J}_\psi/b \sim \mathcal{E}-2\,\mathcal{J}_\psi/b \rightarrow 0$. By expanding (\ref{dispersion relation GM}) we find the following quadratic relation
\begin{equation}
\label{GM dispersion in BMN limit}
\left(\mathcal{E}-\frac{2-b}{b}\,\mathcal{J}_\psi-\mathcal{J}_\phi \right)^2
\,=\,\big(\mathcal{J}_\psi-\mathcal{J}_\phi \big)^2
+ b^2 \Delta\varphi^2\;.
\end{equation}
Notice that it can be obtained also from (\ref{alpha4 small}). In the special case of the conifold (i.e. for $b=2/3$), it becomes
\begin{equation}
\label{BMN conifold}
\big(\mathcal{E}-2\,\mathcal{J}_\psi-\mathcal{J}_\phi \big)^2
\,=\,\big(\mathcal{J}_\psi-\mathcal{J}_\phi \big)^2
+ \frac{4}{9}\, \Delta\varphi^2
\end{equation}
In the pp-wave limit of $AdS_5 \times T^{1,1}$ \cite{Itzhaki:2002kh, Gomis:2002km, Pando Zayas:2002rx}\footnote{See also \cite{Berenstein:2007wi, Berenstein:2007kq} for more recent results.}, the combination\footnote{Here $J^Z_1$ and $J^Z_2$ occur in $H$ with opposite sign with respect to e.g. \cite{Itzhaki:2002kh} because of our opposite choice for the signs of $\phi_1$ and $\phi_2$ in the metric (\ref{metricT11}).}  of charges
$H \equiv \Delta-J_R/2-J^Z_1-J^Z_2$ has been introduced to classify the string states.
By identifying $\Delta=E$, $J_R/2=J_\psi$, $J_1^Z=J_{\phi_1}$ and $J_2^Z=J_{\phi_2}$, and using the fact that for $(\theta_2,\phi_2)=(\pi,\textrm{const})$ we have $J_{\psi}=J_{\phi_2}$ (see (\ref{Jpsi=J2})), we find that in the sector we are considering $H=E-2J_\psi-J_\phi$, which is the combination occurring in l.h.s. (\ref{BMN conifold}).\\
From (\ref{GM dispersion in BMN limit}), one observes that it is possible either to rescale both $H$ and $\mathcal{J}_\psi-\mathcal{J}_\phi$ by the same factor or to rescale $\Delta\varphi$ in order to get the same expression occurring in this limit for the sphere among the corresponding quantities.\\
{\bf The GKP limit.} The GKP regime corresponds to $v = 0$. From (\ref{gamma}) and (\ref{R}), it is easy to check that in this case $R + \gamma = - \cos\beta = 1$, and therefore $\Delta \varphi = \beta =  \pi$, as expected. Moreover, from (\ref{Delta theta b}) we find $\Delta \theta = \pi$ in this limit. The general dispersion relation (\ref{dispersion relation GM}) reduces to
\begin{equation}
\label{dispersion relation gkp}
\mathcal{E}-2\,\mathcal{J}_\phi/b\,=\, 2\,\sqrt{1-b}\,
\cot\left(\frac{\sqrt{1-b}}{2 \,b}\,(\mathcal{E}-2\,\mathcal{J}_\psi/b)\right)\;.
\end{equation}
Thus, while in the pointlike string regime the dispersion relation becomes quadratic, in the folded string limit it remains transcendantal.

\subsection{The limit of  ``basic'' giant magnons}
\label{subsection special cases}

In this subsection we consider two special limits for the giant magnon solution found above: the case of the ``basic'' giant magnons, which is characterized by $\,\mathcal{J}_\psi = \mathcal{J}_\phi$, and the case of $\Omega = (2-b)/b$ (see (\ref{ratio 0})).

\begin{figure}[h]
\begin{tabular}{ccc}
\includegraphics[width=7.3cm]{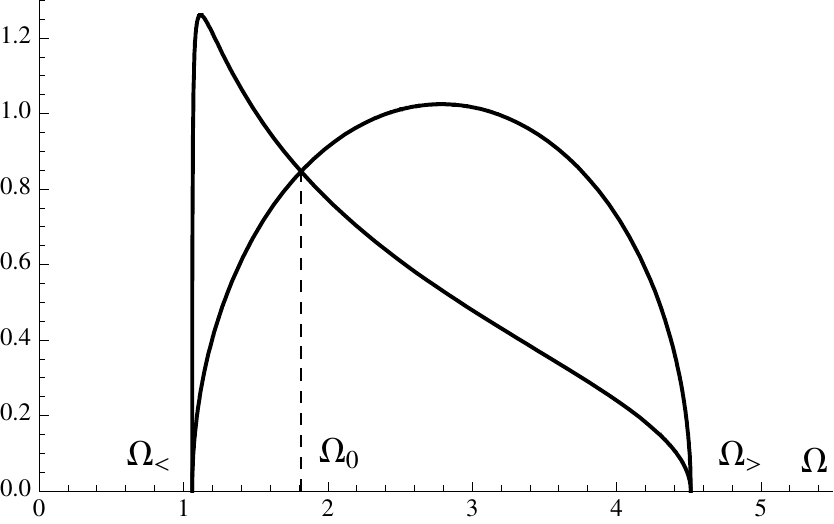}
& & 
\includegraphics[width=7.3cm]{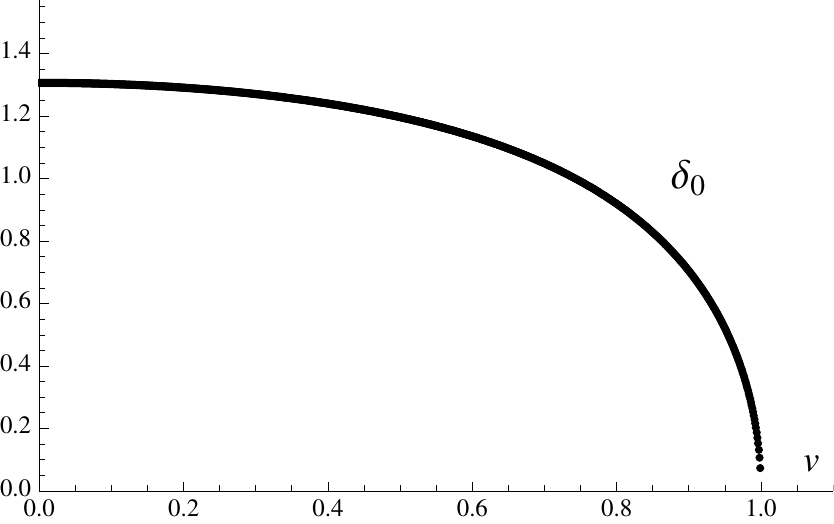}
\end{tabular}
\caption{The basic giant magnons for the conifold (i.e. $b=2/3$). On the left, the graphic solution of the transcendental equation (\ref{trans eq}) for $v=0.89$. On the right, the plot of $\delta_0(\Omega_0(v),v)$, from (\ref{trans eq}).
\label{Jpsi=Jphi pictures}}
\end{figure}

\noindent
{\bf The basic giant magnons.} This limit is characterized by $\,\mathcal{J}_\psi = \mathcal{J}_\phi$ and this equality (see (\ref{GM charges})) means
\begin{equation}
\label{trans eq}
\frac{b\,\delta(\Omega, v)}{\sqrt{1-b}}=\sqrt{\hat{\alpha}_4 (\Omega,v)}
\end{equation}
where $\hat{\alpha}_4(\Omega,v)$ is given in (\ref{alpha4 common}), while $\delta(\Omega,v)$ comes from (\ref{delta def}) and (\ref{R}). The equation (\ref{trans eq}) provides $\Omega=\Omega(v)$ of the basic giant magnons, which will be denoted by $\Omega_0(v)$.\\
The equation (\ref{trans eq}) is transcendental in terms of $v$ and $\Omega$, therefore we will solve it numerically. Plotting the two sides of (\ref{trans eq}), one observes that  a unique solution exists, for each value of $v\in [0,1]$ (for instance, see the picture on the left of the figure \ref{Jpsi=Jphi pictures}). In this way one finds the function  $\Omega_0(v)$ solving (\ref{trans eq}), which is given by the black line in the picture on the right of the figure \ref{zoom bound omega picture}.
Once $\Omega_0(v)$ is known, all the other quantities become functions of $v$ only (for instance, see the picture on the right of the figure \ref{Jpsi=Jphi pictures}, where $\delta_0(v)=\delta(\Omega_0(v),v)$ is plotted in the case of the conifold). 
\begin{figure}[h]
\begin{tabular}{ccc}
\includegraphics[width=7.3cm]{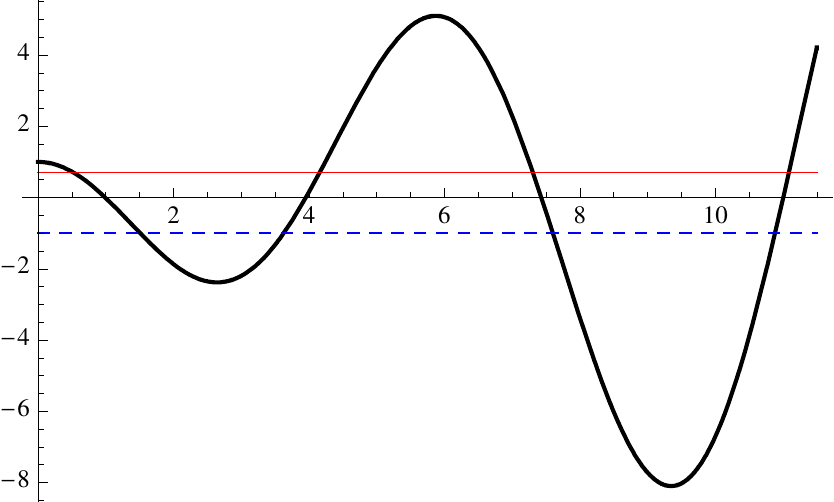}
& & 
\includegraphics[width=7.3cm]{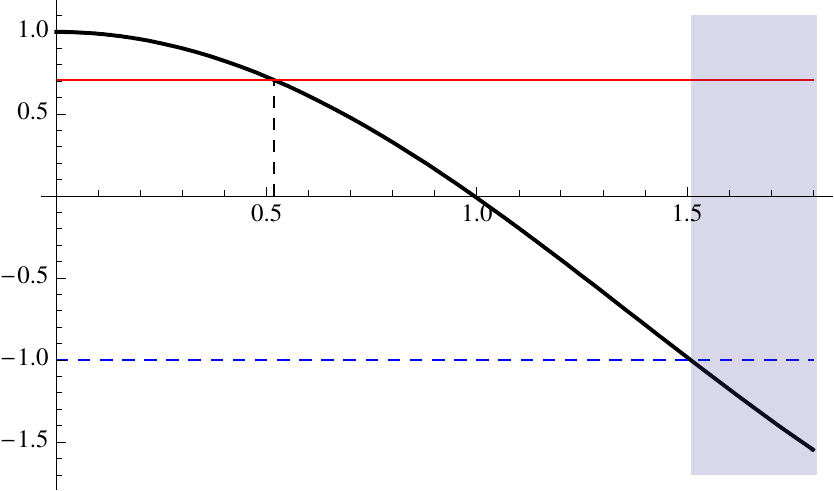}
\end{tabular}
\caption{Inversion of the dispersion relation for basic giant magnons (\ref{dispersion relation Jpsi=Jphi}) on the conifold (i.e. $b=2/3$). The black curve is the l.h.s. of (\ref{dispersion relation Jpsi=Jphi}) as a function of $(\mathcal{E}-2\,\mathcal{J}/b)|_{b=2/3}$, while the horizontal lines give the r.h.s. of (\ref{dispersion relation Jpsi=Jphi}) for a fixed value of $\Delta\varphi \in [0,\pi]$ (in the pictures the red line corresponds to $\Delta\varphi = \pi/4$) and the blue one to $\Delta\varphi=\pi$. On the left, one sees that, for $(\mathcal{E}-2\,\mathcal{J}/b)|_{b=2/3} \geqslant 0$, many solutions to (\ref{dispersion relation Jpsi=Jphi}) exist for a fixed value of $\Delta\varphi$. On the right, the restriction of the same plot to the domain of $(\mathcal{E}-2\,\mathcal{J}/b)|_{b=2/3}$ where (\ref{dispersion relation Jpsi=Jphi}) is invertible and such that $(\mathcal{E}-2\,\mathcal{J}/b)|_{b=2/3} \rightarrow 0$ when $\Delta\varphi \rightarrow 0$ (BMN limit).
\label{basic GM dispersion pictures}}
\end{figure}

\begin{figure}[h]
\begin{center}
\includegraphics[width=7.8cm]{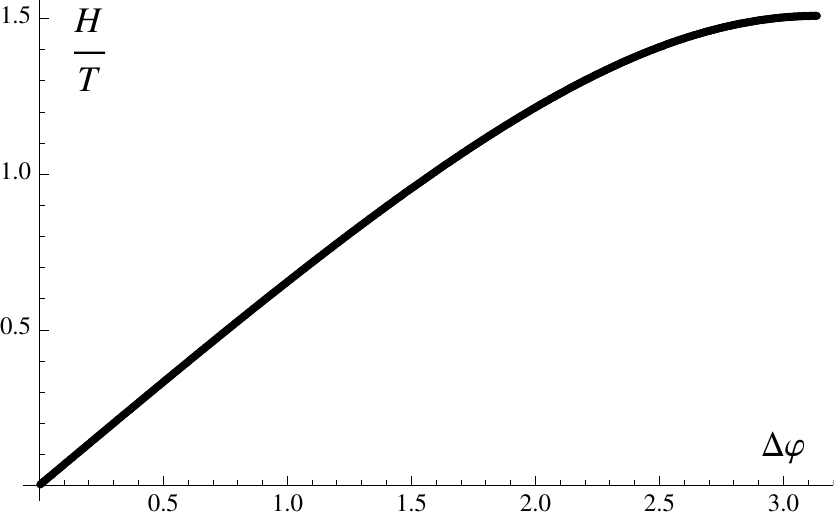}
\end{center}
\caption{The dispersion relation for the basic giant magnons on the conifold. This plot is very similar to the corresponding one for the giant magnons on $\mathbb{R} \times S^2$, found in \cite{Hofman:2006xt}.
\label{disp rel basic GM}}
\end{figure}

\noindent The dispersion relation for the basic giant magnons is found by setting $\mathcal{J}_\psi=\mathcal{J}_\phi\equiv \mathcal{J}$ into (\ref{dispersion relation GM}). The resulting relation can be written as
\begin{equation}
\label{dispersion relation Jpsi=Jphi}
\cos\left(\frac{\sqrt{1-b}}{b}\,(\mathcal{E}-2\,\mathcal{J}/b)\right) -
\frac{\mathcal{E}-2\,\mathcal{J}/b}{2\,\sqrt{1-b}}\,
\sin\left(\frac{\sqrt{1-b}}{b}\,(\mathcal{E}-2\,\mathcal{J}/b)\right)
\,=\,
\cos \Delta \varphi
\end{equation}
This is still a transcendental equation, therefore we can solve it in a graphical way, as done for (\ref{trans eq}), in order to plot $\mathcal{E}-2\,\mathcal{J}/b$ in terms of $ \Delta \varphi$. By plotting the two sides of (\ref{dispersion relation Jpsi=Jphi}) in order to find $\mathcal{E}-2\,\mathcal{J}/b \geqslant 0$ in terms of $\Delta \varphi \in [0,1]$, one immediately observes that the equation (\ref{dispersion relation Jpsi=Jphi}) has many solutions for a fixed $b \in (0,1)$ (see the picture on the left of the figure \ref{basic GM dispersion pictures}). For $\Delta \varphi \in [0,\pi]$, there are infinite disconnected intervals where (\ref{dispersion relation Jpsi=Jphi}) can be inverted.
To find the right one, we require that $(\mathcal{E}-2\,\mathcal{J}/b) \rightarrow 0$ when $\Delta\varphi \rightarrow 0$ (BMN limit, see the subsection \ref{BMN and GKP}). Thus, in the domain $[0,(\mathcal{E}-2\,\mathcal{J}/b)_{\textrm{max}}]$, where $(\mathcal{E}-2\,\mathcal{J}/b)_{\textrm{max}}$ corresponds to $\Delta\varphi=\pi$, we can invert (\ref{dispersion relation Jpsi=Jphi}) and find $\mathcal{E}-2\,\mathcal{J}/b$ for any $\Delta\varphi \in [0,\pi]$ (see the picture on the right of the figure \ref{basic GM dispersion pictures}). The analytic expression of $H/T=(\mathcal{E}-2\,\mathcal{J}/b)|_{b=2/3}$ in terms of $\Delta\varphi$ cannot be found, but its plot is given in the figure \ref{disp rel basic GM} and it resembles the corresponding plot for the giant magnons on $\mathbb{R} \times S^2$.

\noindent
It is interesting to consider the basic giant magnon in the limit of the three dimensional sphere.
In the limit $b \rightarrow 1$, the transcendental equation (\ref{trans eq}) becomes
\begin{equation}
\label{transEq b=1}
\frac{\sqrt{4\,\Omega_0|_{b=1}-v^2(1+\Omega_0|_{b=1})^2}}{\Omega_0|_{b=1}}
\,=\,
\sqrt{4\,\Omega_0|_{b=1}-v^2(1+\Omega_0|_{b=1})^2}\;.
\end{equation}
Since $\hat{\alpha}_4|_{b=1}=4\,\Omega-v^2(1+\Omega)^2 >0$, the solution of (\ref{transEq b=1}) is $\Omega_0|_{b=1}=1$ for any $v \in [0,1]$; therefore\footnote{We recall that $\Omega|_{b=1}=(\omega_1+\omega_2)/(\omega_1-\omega_2)$.} $\omega_2=0$, which implies $\Phi_2'=0$, as one can check from (\ref{Psiprime 1}) and (\ref{Phiprime 1}). This corresponds to consider $S^2 \subset S^3$. Indeed, for the basic giant magnons, when $b \rightarrow 1$ we have $\mathcal{J}_1=2\,\mathcal{J}|_{b=1}$ and $\mathcal{J}_2=0$, and (\ref{dispersion relation Jpsi=Jphi}) in this limit gives
\begin{equation}
\label{dispersion relation S2}
(\mathcal{E}-2\,\mathcal{J})^2
\,=\,
4 \sin^2\frac{\Delta \varphi_1}{2}
\end{equation}
which is the dispersion relation for $\mathbb{R}\times S^2$ found in \cite{Hofman:2006xt}.\\

\noindent 
{\bf The ratio $\Omega=(2-b)/b$.}
To close the discussion about the giant magnons, we briefly consider also the special value (\ref{ratio 0}) for $\Omega$, which corresponds to the horizontal green line in the picture on the right in the figure \ref{picture potential GM} (right) and in the figure \ref{zoom bound omega picture}. \\
From (\ref{Psiprime 1}) and (\ref{Phiprime 1}) specialized to the giant magnon solution (i.e. with $A_\phi$ given by (\ref{AGM})), we can see that for the ratio (\ref{ratio 0}), we get finite $\Delta\psi$ and $\Delta\phi$, and also
\begin{equation}
\label{angular amplitude special ratio}
\Delta\psi \,=\, \Delta\phi\,=\,\Delta\varphi\;.
\end{equation}
Notice that, for every choice of $\Omega$ constant in $v$ and $\Omega \neq (2-b)/b$, an upper bound $v_{\textrm{max}}<1$ exists on the values of $v$ allowed by the condition $\alpha_4>0$. Instead, both the cases of $\Omega=\Omega(v)$ we have considered in this section are defined for any $v \in [0,1]$.

\section{Single spike strings: energy and momenta}
\label{section dispersion relation SS}

In this section we compute the conserved charges for the single spike strings, as done in the previous section for the giant magnons.

The single spike strings are given by the solution (\ref{solution u}) with (\ref{ASS}) for $A_\phi$.
Only $\alpha_4$ and $k$ change with respect to the case of the giant magnons. Taking the positive root for $k$  in (\ref{k2SS}), now the three momenta in (\ref{pt}), (\ref{ppsi}) and (\ref{pphi}) read respectively
\begin{eqnarray}
\label{ptSS}
p_t &=&-\,T \; \frac{b}{2\,d}\,(\omega_\psi+\omega_\phi)
\\
\label{ppsiSS}
\rule{0pt}{.8cm}
\frac{p_\psi}{b} 
&=& \frac{T\,b\,c\,\omega_\phi}{2\,d^2(1-v^2)}\, u(y) \\
\label{pphiSS}
\rule{0pt}{.8cm}
\frac{p_\phi}{b} 
&=& \frac{T\,c\,\omega_\phi}{2\,d^2(1-v^2)}\, \Big( 2\,(1-b)\,u(y)^2
+\big(b\,\Omega-2\,(1-b)\big)\,u(y)\Big)
\phantom{xxxxx}
\end{eqnarray}
where $u(y)$ is given by (\ref{solution u}) and $v=c/d<1$. Notice that now, among the three conserved charges $\mathcal{E}$, $\mathcal{J}_\psi$ and $\mathcal{J}_\phi$, the energy $\mathcal{E}$ is the only divergent one. Moreover, for the finite expressions of  $\mathcal{J}_\psi$ and $\mathcal{J}_\phi$ of the single spike strings we find that
\begin{equation}
\label{SS charges}
\frac{2}{b}\,\mathcal{J}_\psi
\,=\,
\frac{b\,\delta}{\sqrt{1-b}}
\hspace{3cm}
\frac{2}{b}\,\mathcal{J}_\phi
\,=\,
 \sqrt{\hat{\alpha}_4}\;.
\end{equation}
Differently from the giant magnons, now $\Delta \varphi$ diverges, telling us that the single spike strings wind infinitely many times along the direction of $\varphi$. We can construct a finite quantity as a linear combination of the divergent ones as follows
\begin{equation}
\label{SS finite combination}
\frac{\mathcal{E}}{b}\,- \Delta\varphi
 \,=\, \beta
\end{equation}
where (\ref{Utilde}) has been used. We remark that $2\,\mathcal{J}_\psi/b$, $2\,\mathcal{J}_\phi/b$ and $\mathcal{E}/b-\Delta\varphi$ for the single spike strings play the same role respectively of $\mathcal{E}-2\,\mathcal{J}_\psi/b$, $\mathcal{E}-2\,\mathcal{J}_\phi/b$ and $\Delta\varphi$ for the giant magnons, as can be seen by comparing (\ref{GM charges}) with (\ref{SS charges}) and (\ref{DeltaPhib GM}) with (\ref{SS finite combination}). 
Thus, from (\ref{alpha4 relation 1}) we get the following dispersion relation for the single spike strings
\begin{equation}
\label{dispersion relation SS}
\frac{\mathcal{J}_\phi}{b\,\sqrt{1-b}}
\,=\,
\frac{\cos\big(2\sqrt{1-b}\,\mathcal{J}_\psi/b^2)-\cos (\mathcal{E}/b-\Delta\varphi)}{\sin(2\sqrt{1-b}\,\mathcal{J}_\psi/b^2)}\;.
\end{equation}
As done before, we check this expression by showing that it reproduces the known result on the three dimensional sphere when $b \rightarrow 1$. In this limit, (\ref{dispersion relation SS}) gives\footnote{We stress that $\mathcal{J}_\psi|_{b=1}=(\mathcal{J}_1+\mathcal{J}_2)/2$ and $\mathcal{J}_\phi|_{b=1}=(\mathcal{J}_1-\mathcal{J}_2)/2$.}
\begin{equation}
\label{dispersion relation S3 from SS}
\mathcal{J}_\psi\,\mathcal{J}_\phi
\,=\,
\sin^2\left(\frac{\mathcal{E}-\Delta \varphi_1}{2}\right)
\end{equation}
which is dispersion relation of single spike strings on $\mathbb{R}\times S^3$ found in \cite{Ishizeki:2007we}.

\section{Long gauge theory operators}
\label{section gauge theory}

In this section we discuss in a qualitative way the gauge theory operators dual to the giant magnon solution we have studied in the previous sections. We hope that quantitative checks through gauge theory computations will be done in the future.

The gauge theory dual to type IIB on $AdS_5 \times T^{1,1}$ was found by Klebanov and Witten in \cite{Klebanov:1998hh}. It is a $\mathcal{N}=1$ supersymmetric theory whose gauge group is $SU(N)\times SU(N)$ and with bifundamental chiral superfields $A_i$, $B_j$ ($i,j=1,2$) transforming respectively in the $(N,\bar{N})$ and $(\bar{N},N)$ representations of the gauge group.
The superpotential reads
\begin{equation}
W\,=\,\lambda_c\,\epsilon^{ij} \epsilon^{kl} \,\textrm{tr}(A_i B_k A_j B_l)
\end{equation}
where $\lambda_c$ is the coupling constant. The global symmetries are $SU(2)\times SU(2)\times U(1)_R \times U(1)_B$, where the $SU(2)$'s act on $A_i$ and $B_j$ respectively, $U(1)_R$ is the anomaly free $R$ symmetry and $U(1)_B$ is the baryonic symmetry. Under the $R$ symmetry $A_i$ and $B_j$ have the same charge $1/2$, while under the baryonic symmetry they have opposite charges.
As CFT primary operators, these fundamental fields have dimensions $\Delta=3/4$. \\
Hereafter we denote by $A_i$ and $B_k$ the scalar components of the chiral superfields just described
and by $\bar{A}_i$ and $\bar{B}_k$ their complex conjugate fields. 
The quantum numbers characterizing them are the conformal dimensions $\Delta$, the $R$ charge $J_R$ and the quantum numbers $J^Z_1$ and $J_2^Z$ for the two  global $SU(2)$'s\footnote{We refer to tables in \cite{Itzhaki:2002kh} for a complete list of the fields with the corresponding dimensions and charges. We stress that our definition of $H$ is different from  \cite{Itzhaki:2002kh} because we have different signs for  $\phi_1$ and $\phi_2$ in the metric.}.

Strings moving on $T^{1,1}$ should be dual to pure scalar operators of the gauge theory, not containing fermions, covariant derivatives or gauge field strengths. Moreover, our string solutions has large energy $E$ and spin $J_\psi$, which means large $\Delta$ and large $J_R$  for the dual gauge operators. Thus, the gauge invariant operators we consider are of the following form
\begin{equation}
\label{long operator}
\textrm{tr}(\dots AB \dots \bar{B}\bar{A} \dots A\bar{A} \dots\bar{B} B \dots)
\end{equation}
and they are long, i.e. there are infinitely many pairs within the trace. Each of the four combinations in (\ref{long operator}) represents a closed loop on the quiver diagram of the dual gauge theory. In the spin chain language, the operator  (\ref{long operator}) is associated to a spin chain state and each combination representing a closed loop on the quiver diagram is associated to a site of the chain in a certain spin state. Thus, since we have four possibilities for each combination in (\ref{long operator}), there are sixteen choices for each site.\\
As already mentioned, the quantum numbers of the gauge theory are identified with the conserved charges of the string theory as follows: $\Delta = E$, $J_R/2=J_\psi$, $J^Z_1=J_{\phi_1}$ and $J^Z_2=J_{\phi_2}$. Moreover, in the sector of the conifold we have considered we have $J_\psi=J_{\phi_2}$ (see the appendix \ref{appendix symmetries}), therefore we look for gauge operators having $J_R/2=J^Z_2$.\\
First, we consider the following inequality between $\Delta$ and $J_R$
\begin{equation}
\label{bound}
\Delta\,\geqslant\,\frac{3}{2}\,|J_R|
\end{equation}
which comes from the unitarity bound of the $\mathcal{N}=1$ superconformal algebra\footnote{See e.g. \cite{Kim:2003vn}, where $J_R$ denotes what we call $J_\psi$.}. Once written in terms of the string variables $E$ and $J_\psi$, the inequality (\ref{bound}) is satisfied by both the giant magnons ($\delta \geqslant 0$ in the first equation of (\ref{GM charges}) with $b=2/3$) and the single spike strings (this case is trivial, since $E$ is divergent while $J_\psi$ is finite).\\
Let us consider the following BPS operator
\begin{equation}
\label{ground state}
\textrm{tr}(A_1 B_1)^L
\end{equation}
with large $L$. Among the operators of the form (\ref{long operator}), it is the only one having $H=0$, therefore we take it as the ``ground state''\footnote{If one instead considers the metric with the signs as in \cite{Itzhaki:2002kh}, then the ground state is $\textrm{tr}(A_2 B_2)^L$ and the corresponding geodesic is given by $\theta_1=\theta_2=0$ (see also \cite{Kim:2003vn}).}.
Moreover, it satisfies $J_R/2 = J^Z_1= J^Z_2$. \\
Then, we consider the following long operators
\begin{equation}
\label{excitation}
\textrm{tr}\big(X(A_1 B_1)^L\big)
\end{equation}
obtained by inserting the operator $X$ within the trace of (\ref{ground state}). Taking for $X$ one of the sixteen operators discussed above (of course it is assumed $X \neq A_1 B_1$), we find that we have only five ways to get $H=1$.
The relevant bosonic operators in the BMN regime which do not involve derivatives are $A_1B_2$, $A_2B_1$, $A_1\bar{A}_2$ and $\bar{B}_2B_1$ \cite{Itzhaki:2002kh, Gomis:2002km, Pando Zayas:2002rx}\footnote{Also $X=\bar{B}_2 \bar{A}_2$ has $H=1$.}, which are shown in the table (\ref{excited states}).
Notice that $A_1\bar{A}_2$ and $\bar{B}_2B_1$ have $\Delta = 2$, namely the free field value, because they belong to the supermultiplet of the currents.
The remaining 10 choices have $H>1$.
Among the four operators with $H=1$ in (\ref{excited states}), only $A_2B_1$ and $A_1\bar{A}_2$ have $J_R/2 = J^Z_2$. Thus, if we want to excite the ground state (\ref{ground state}) with an $H=1$ operator mantaining the condition $J_R/2 = J^Z_2$ for any $L$, then we have to choose between $X=A_2B_1$ and $X=A_1\bar{A}_2$. Instead, if we require $J_R/2 = J^Z_2$ for (\ref{excitation}) only at large $L$, then all the five $X$'s with $H=1$ can be used.
\begin{equation}
\label{excited states}
\begin{array}{|c|c|c|c|c|c|}
\hline
X & \Delta & J_R & J^Z_1 & J_2^Z & H \equiv \Delta-J_R/2-J_1^Z-J_2^Z\\
\hline
A_1B_1 & 3/2 & 1 & 1/2 & 1/2 & 0 \\
\hline
A_1B_2 & 3/2 & 1 & 1/2 & -1/2 & 1 \\
\hline
A_2B_1 & 3/2 & 1 & -1/2 & 1/2 & 1 \\
\hline
A_1\bar{A}_2 & 2 & 0 & 1 & 0 & 1 \\
\hline
\bar{B}_2B_1 & 2 & 0 & 0 & 1 & 1 \\
\hline
\end{array}
\end{equation}
Given that $J_R/2 = J^Z_2$ in the sector we are considering, for the gauge operator dual to the giant magnons is \cite{Itzhaki:2002kh, Gomis:2002km, Pando Zayas:2002rx}
 \begin{equation}
 O_p\,=\,
 \sum_l e^{i p l}\,
\big(\,\dots (A_1 B_1)_{l-1}\, X_l \,(A_1 B_1)_{l+1} \dots \,\big)
\end{equation}
where $X$ is either $A_2 B_1$ or $A_1\bar{A}_2$ and $p$ is the momentum of the impurity, located at the $l$th site of the spin chain. 
The fact that we have these two candidates is expected, since it is possible to act on our giant magnon solutions with a residual $SO(2)$ global symmetry. This is analogous to the presence of two possible impurities $Y$ and $\bar{Y}$ on the spin chain\footnote{We take $\textrm{tr}(\dots ZZZZ\dots)$ as the ground state.} describing the $SO(4)$ sector of $\mathcal{N}=4$ SYM.

\section{Conclusions}

In this paper we have studied the giant magnons and the single spike strings on the squashed three dimensional sphere, which gives a sector of the conifold for a special value of the squashing parameter. The finite combinations of the global charges occur in the dispersion relation in a transcendental way, 
which suggests that the integrable structure on the conifold, if it exists, is more complicated than the one on the sphere.\\
There are many interesting ways to go beyond our work. For instance, on the string theory side, one can study the giant magnons in our sector with finite size\footnote{The relation with the compound KdV discussed in the appendix \ref{appendix compound KdV} can be useful for this analysis.} or again with infinite size but on the whole $\mathbb{R} \times T^{1,1}$.\\
The possible presence of an integrable structure underlying the type IIB string theory on $AdS_5 \times T^{1,1}$ and its dual gauge theory is an important issue which is still unclear and deserves further studies.

\subsection*{Acknowledgments}

We are grateful to Marcus Benna and Igor Klebanov for collaboration at the initial stage of this project. It is a pleasure to thank Gleb Arutyunov, Pasquale Calabrese, Davide Fioravanti, Giuseppe Mussardo, Michele Papucci and Andrew Strominger for discussions. We are grateful in particular to Igor Klebanov for the having read the draft and for many useful discussions and comments.
E.T. acknowledges the Physics departments of Bologna university and Princeton university for the kind hospitality during parts of this work.\\
S.B. is supported in part by the National Science Foundation under Grant No. PHY-0756966.

\appendix

\section{A sector of the conifold}
\label{appendix sector T11}

In this appendix we show that 
the submanifold characterized by $(\theta_2,\phi_2)= \textrm{constant}$ (or $(\theta_1,\phi_1)= \textrm{constant}$ as well) is a consistent sector.\\
Given the ten dimensional target space metric  $G_{MN}$, the Polyakov action in the conformal gauge is\footnote{We recall that the string tension $T$ is related to the 't Hooft coupling $\lambda \equiv g_{\scriptscriptstyle \textrm{YM}}^2 N$ as $T=\frac{\sqrt{\lambda}}{2\pi}$.}
\begin{equation}
\label{polyakov action}
S\,=\,\int d\tau\,d\sigma\;\mathcal{L}
\,=\,
-\,\frac{T}{2}\,\int d\tau\,d\sigma\;
G_{MN}\left(-\,\partial_\tau X^M \partial_\tau X^N + \,\partial_\sigma X^M \partial_\sigma X^N\,\right)\;.
\end{equation}  
It is supported by the Virasoro constraints, which read
\begin{eqnarray}
\label{VC1}
G_{MN}\left(\,\partial_\tau X^M \partial_\tau X^N + \,\partial_\sigma X^M \partial_\sigma X^N \,\right) &=&0
\\
\label{VC2}
G_{MN} \,\partial_\tau X^M \partial_\sigma X^N &=& 0\;.
\end{eqnarray} 
The momentum $p_M$ canonically conjugate to the coordinate $X^M$ is
\begin{equation}
\label{momenta}
p_M\,=\,\frac{\partial \mathcal{L}}{\partial \dot{X}^M}\,=\,T\,G_{MN}\,\partial_\tau X^N
\end{equation} 
where $ \dot{X}^M \equiv \partial_\tau X^M$.
In order to simplify the problem, we want to restrict our analysis to a consistent sector of the ten dimensional space. 
A submanifold is a consistent sector of the theory when the differential equations of the classical theory (equations of motion and Virasoro constraints) obtained by restricting the ones of the full space to the submanifold are the same ones obtained by taking the submanifold as target space.
For instance, from the ten equations of motion coming from (\ref{polyakov action}) and the two Virasoro constraints (\ref{VC1}) and (\ref{VC2}), it is not difficult to realize that, when the target space metric is factorized like $AdS_5 \times M^5$, a consistent sector is given by the submanifold $\mathbb{R}\times M^5$ obtained by setting to constants all the coordinates of $AdS_5$ except for the time coordinate $t \in \mathbb{R}$.
\\
In this paper we consider the case of the conifold, namely $M^5 = T^{1,1}$; therefore the target space we start with is  $\mathbb{R}\times T^{1,1}$, whose metric reads
\begin{equation}
\label{RtimesT11}
ds^2\,=\,
-\,dt^2+
\frac{1}{9}\big(d\psi- \cos\theta_1 d\phi_1- \cos\theta_2 d\phi_2\big)^2
+ \frac{1}{6}\big(d\theta_1^2+\sin^2\theta_1 d\phi_1^2
+ d\theta_2^2+\sin^2\theta_2 d\phi_2^2\big)
\end{equation}
where $0\leqslant \psi < 4\pi$ parameterizes the $U(1)$ fiber,  $0\leqslant \theta_i < \pi$, 
$0\leqslant \phi_i < 2\pi$ ($i=1,2$)  describe two $S^2$'s and $t\in \mathbb{R}$ is the time coordinate of $AdS_5$. \\
In order to find a consistent sector of (\ref{RtimesT11}), first we write the equations of motion coming from (\ref{RtimesT11}).
Beside the simple equation $\partial_\mu \partial^\mu \,t \equiv - \,\partial^2_\tau\, t + \partial^2_\sigma \,t =0$ for the time coordinate $t$, the ones coming from the remaining angular coordinates are
\begin{eqnarray}
\label{eqPsi}
& &\hspace{-.5cm}  
\partial_\mu \big(\partial^\mu \psi 
- \cos\theta_1\,\partial^\mu\phi_1
- \cos\theta_2\,\partial^\mu\phi_2
\big)\, =\, 0\\
\rule{0pt}{.8cm} 
\label{eqPhi}
& &\hspace{-.5cm}  
\partial_\mu \left(\,
\frac{1}{9}\,\big(
\cos\theta_i\,\partial^\mu \psi 
- \cos\theta_1 \cos\theta_2\,\partial^\mu \phi_j \big)
- \left(\,\frac{1}{9} \cos^2\theta_i+\frac{1}{6} \sin^2\theta_i\right) \partial^\mu \phi_i \right) =\, 0
\phantom{xxxxx}\\
& & \hspace{-.5cm}  
\rule{0pt}{.7cm} 
\frac{1}{3}\,\sin\theta_i\, \Big(
2  \big(\partial_\mu \psi - \cos\theta_j\,\partial_\mu \phi_j\big) + \cos\theta_i\,\partial_\mu \phi_i\Big)
\,\partial^\mu \phi_i
- \partial_\mu \partial^\mu  \theta_i \,=\,0
\end{eqnarray}
where the indices $i\neq j$ take the values 1 or 2.\\
Now, setting the fields $\phi_2(\tau,\sigma)$ and $\theta_2(\tau,\sigma)$ to constant values, we are left with the following equations\footnote{Notice that (\ref{eqPhi}) for $i=2$ and (\ref{eqPsi}) reduce to the same equation (\ref{eqPsiRed}) for $(\theta_2,\phi_2)= \textrm{constant}$.}
\begin{eqnarray}
\label{eqPsiRed}
& &\hspace{-.5cm}    
\partial_\mu \big(\partial^\mu \psi 
- \cos\theta_1\,\partial^\mu\phi_1
\big)\, =\, 0\\
\rule{0pt}{.7cm} 
\label{eqPhiRed}
& &\hspace{-.5cm}  
\rule{0pt}{.8cm} 
\partial_\mu \left(\,
\frac{1}{9}\,\cos\theta_1\,\partial^\mu \psi 
- \left(\,\frac{1}{9} \cos^2\theta_1+\frac{1}{6} \sin^2\theta_1\right) \partial^\mu \phi_1 
\right)=\, 0
\phantom{xxxxx}\\
& &\hspace{-.5cm}  
\rule{0pt}{.7cm} 
\frac{1}{3}\,\sin\theta_1\, \big(
2  \, \partial_\mu \psi + \cos\theta_1\,\partial_\mu \phi_1\big)
\partial^\mu \phi_1
- \partial_\mu \partial^\mu  \theta_1 \,=\,0\;.
\end{eqnarray}
Now, these equations can be found also as the equations of motion coming from the Polyakov action (\ref{polyakov action}) equipped with a target space metric (\ref{RtimesT11}) reduced to the submanifold with constant $(\theta_2,\phi_2)$, namely
\begin{equation}
\label{RtimesT11reduced}
ds^2\,=\,
-\,dt^2+
\frac{1}{9}\big(d\psi- \cos\theta_1 d\phi_1\big)^2
+ \frac{1}{6}\big(d\theta_1^2+\sin^2\theta_1 d\phi_1^2\big)\;.
\end{equation}
Similarly, one can check that this reduction works also for the Virasoro constraints.
Thus, the submanifold characterized by $(\theta_2,\phi_2)= \textrm{constant}$ is a consistent sector and, symmetrically, the same can be said for $(\theta_1,\phi_1)=\textrm{const}$. 
We stress that not all the submanifolds are consistent sectors. For instance, 
the submanifold $(\phi_1,\phi_2)=\textrm{const}$ is not a consistent sector because the equations (\ref{eqPhi}) reduced  to this submanifold provide also the non trivial equations $\partial_\mu (
\cos\theta_i\,\partial^\mu \psi)=0$, which cannot be obtained as equations of motion of the Polyakov action with the reduced metric.

\section{Symmetries and momenta of $T^{1,1}$}
\label{appendix symmetries}

In this appendix we discuss the infinitesimal trasformations which leave the metric of the conifold invariant up to first order included in the infinitesimal parameters occurring in the transformations.

The metric on $T^{1,1}$ is
\begin{equation}
\label{metricT11}
ds^2_{T^{1,1}}\,=\,
\frac{1}{9}\big(d\psi- \cos\theta_1 d\phi_1- \cos\theta_2 d\phi_2\big)^2
+ \frac{1}{6}\big(d\theta_1^2+\sin^2\theta_1 d\phi_1^2
+ d\theta_2^2+\sin^2\theta_2 d\phi_2^2\big)\;.
\end{equation}
Given the following infinitesimal transformations
\begin{eqnarray}
 \delta \theta_i &=& -\,\varepsilon^X_i  \sin(\phi_i)  + \varepsilon^Y_i \cos(\phi_i) \\
 \rule{0pt}{.8cm}
 \delta \phi_i  &=&  -\,\varepsilon^X_i  \,\frac{\cos(\phi_i)}{\tan(\theta_i)} - \varepsilon^Y_i\, \frac{\sin(\phi_i)}{\tan(\theta_i)} + \varepsilon^Z_i \\
 \rule{0pt}{.8cm}
\delta \psi &=& -\sum_{i=1,2}\left(  \varepsilon^X_i \, \frac{\cos(\phi_i)}{\sin(\theta_i)} + \varepsilon^Y_i\, \frac{\sin(\phi_i)}{\sin(\theta_i)} \right)+ \varepsilon^{\psi}
\end{eqnarray}
where $i=1,2$ and the seven $\varepsilon$'s are the infinitesimal parameters, one can check that they leave (\ref{metricT11}) invariant up to $O(\varepsilon^2)$ terms. Notice that the one form of the fibration and the two metrics of the $S^2$'s are separately invariant up to $O(\varepsilon^2)$ terms.\\
By switching on one of the infinitesimal parameters only, we can find its associated momentum. For $\varepsilon^{\psi}$, $\varepsilon^Z_i$, $\varepsilon^X_i$ and $\varepsilon^Y_i$ we get respectively 
\begin{eqnarray}
p_\psi &=&
\frac{1}{9}\,\big(\partial_\tau\psi-\cos\theta_1\,\partial_\tau\phi_1-\cos\theta_2\,\partial_\tau\phi_2\big)
\\
\rule{0pt}{.8cm}
\label{p_phi_i}
p_{\phi_i} &=&
\frac{1}{6}\,\sin^2\theta_i \,\partial_\tau\phi_i-\cos\theta_i\,p_\psi\\
\rule{0pt}{.8cm}
p_i^X &=&
-\,\frac{1}{6} \,\sin\phi_i \,\partial_\tau\theta_i
-\frac{\cos\phi_i}{\sin\theta_i}\,\big(p_\psi+\cos\theta_i\,p_{\phi_i}\big)\\
\rule{0pt}{.8cm}
p_i^Y &=&
\frac{1}{6} \,\cos\phi_i \,\partial_\tau\theta_i
-\frac{\sin\phi_i}{\sin\theta_i}\,\big(p_\psi+\cos\theta_i\,p_{\phi_i}\big)\;.
\end{eqnarray}
It is useful to consider also
\begin{equation}
\vec{p}_i^{\,\,2}\,=\,(p_i^X)^2+(p_i^Y)^2+(p_{\phi_i} )^2\,=\,
\frac{(\partial_\tau\theta_i)^2}{36}+\frac{(p_\psi+\cos\theta_i\,p_{\phi_i})^2}{\sin^2\theta_i}
+(p_{\phi_i} )^2\;.
\end{equation}
From (\ref{p_phi_i}), we observe that $\theta_i=0$ and $\theta_i=\pi$ give respectively $p_{\phi_i}=-p_\psi$ and $p_{\phi_i}=p_\psi$; while setting $\phi_i$ to constant provides $p_{\phi_i}=- \cos\theta_i\, p_\psi$.
In the sector we are considering we have $\theta_2=\pi$, therefore
\begin{equation}
\label{Jpsi=J2}
p_\psi \,=\, p_{\phi_2}\;.
\end{equation}
Notice that the parameters $\varepsilon^\psi$, $\varepsilon_1^Z$ and $\varepsilon_2^Z$ can also be taken finite. The integral of their associated momenta $p_\psi$, $p_{\phi_1}$ and $p_{\phi_2}$ over $\sigma$ at fixed $\tau$ gives the conserved charges $J_\psi$,  $J_{\phi_1}$ and $J_{\phi_2}$ respectively. The dispersion relation for the giant magnons and for the spiky strings on the full $\mathbb{R} \times T^{1,1}$ will involve, beside the energy $E$, all these conserved charges. 
On the submanifold $(\theta_2,\phi_2)=(\pi,\textrm{const})$ that we are considering, we have $J_\psi=J_{\phi_2}$ (see (\ref{Jpsi=J2})).

\section{The relation with known equations}
\label{appendix known equations}

In this appendix we show how the main differential equation we deal with in our problem is related to the double sine-Gordon equation and to the compound KdV equation.

\subsection{The double sine-Gordon equation}
\label{appendix double sine-gordon}

First we show that, given the ansatz (\ref{GMansatz}), the equation of motion for $\theta(y)$ can be written as a double sine-Gordon equation (see e.g. \cite{Mussardo:2004rw}).
Introducing $\xi(y)$ through $u \equiv \alpha_> \cos^2(\xi/2)$, we have $\xi(y) \in [0,\pi)$ and the differential equation (\ref{total energy GM bis}) can be written as follows
\begin{equation}
\big(\xi'\big)^2\,=\,
-\,4\,\alpha_8\, \alpha_>^2 \cos^2\frac{\xi}{2}\left(\cos^2\frac{\xi}{2}-\frac{\alpha_<}{\alpha_>}\right)\,.
\end{equation}
Then, assuming $\xi' \neq 0$ and deriving this equation w.r.t. $y$, we get
\begin{equation}
\label{double sine-gordon}
\xi''\,=\,
\alpha_8\, \alpha_> (\alpha_>-\alpha_<) \sin\xi+\frac{\alpha_8\,\alpha_>^2}{2}\,\sin(2\xi)
\end{equation}
which is known as double sine-Gordon equation. 
By employing (\ref{positive alpha}) and (\ref{R}), we can write (\ref{double sine-gordon}) also as
\begin{equation}
\label{double sine-gordon 2}
\xi''\,=\,
-\,\frac{2\alpha_4}{1-R}\left( \sin\xi+\frac{1+R}{4}\,\sin(2\xi)\right)\,.
\end{equation}
From this form it is evident that in the case of $S^3$, where $R=-1$, (\ref{double sine-gordon 2}) reduces to the sine-Gordon equation, as expected.

\subsection{The compound KdV equation}
\label{appendix compound KdV}

Now we show how the differential equation (\ref{total energy full}) is related to the compound Korteweg-de Vries (KdV) equation\footnote{We are grateful to Davide Fioravanti for having addressed our attention to the KdV equation and its modifications.}  (see e.g. \cite{Wang96} and references therein).\\
Let us consider the generalized KdV equation
\begin{equation}
\label{generalized KdV}
\widehat{A}\,\partial_\tau u
+\big( \widehat{B}+\widehat{C}\,u^n \big)\,u^n \,\partial_\sigma u
+\widehat{D}\,\partial^3_\sigma u\,=\,0
\end{equation}
where $u=u(\tau,\sigma)$, $n>0$ is an integer number. For $n=1$ and $\widehat{C}=0$, the equation (\ref{generalized KdV}) reduces to the well known KdV equation, while when $n=1$ and $\widehat{C} \neq 0$ it is called compound KdV equation.\\
Introducing the ansatz $u(\tau,\sigma)=u(y)$ where $y=c\,\sigma-d\,\tau$, and integrating once w.r.t. $y$, the equation (\ref{generalized KdV}) becomes
\begin{equation}
\label{generalized KdV step1}
C_1-d\,\widehat{A}\,u+\frac{c\,\widehat{B}}{2}\,u^2+\frac{c\,\widehat{C}}{3}\,u^3
+c^3\,\widehat{D}\,u''
\,=\,0
\end{equation}
where $C_1$ is the integration constant. Then, assuming $u' \neq 0$, we can multiply (\ref{generalized KdV step1}) by $u'$ and integrate again w.r.t. $y$, obtaining 
\begin{equation}
\label{generalized KdV step2}
C_2+ C_1\,u-\frac{d\,\widehat{A}}{2}\,u^2+\frac{c\,\widehat{B}}{6}\,u^3+\frac{c\,\widehat{C}}{12}\,u^4
+c^3\,\widehat{D}\,\frac{(u')^2}{2}
\,=\,0
\end{equation}
where $C_2$ is another integration constant. Now, we observe that the equation (\ref{generalized KdV step2}) is exactly (\ref{total energy full}), after a straightforward identification between the two sets of coefficients.\\
The case of $S^3$ (i.e. $b=1$) corresponds to $\widehat{C}=0$, namely to the KdV equation.

\section{Two remarks on the solution}
\label{appendix 2 remarks}

In this appendix we discuss first the solution of (\ref{total energy GM}) with $\alpha_4 <0$ and then the limiting solution for $\alpha_4 \rightarrow 0$.
While the first case has no relevance for our problem, the second one represents an interesting special case.
\begin{figure}[h]
\begin{tabular}{ccc}
\includegraphics[width=7.3cm]{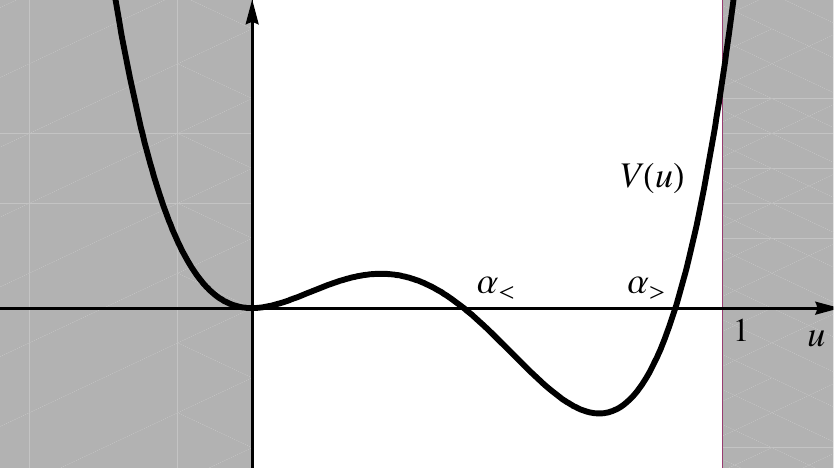}
& & 
\includegraphics[width=7.3cm]{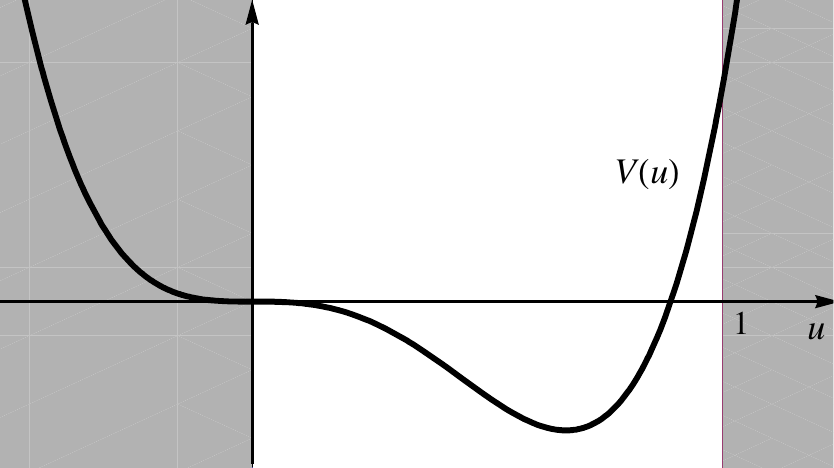}
\end{tabular}
\caption{Special cases of the potential in (\ref{total energy GM}). On the left, the potential  (\ref{total energy GM}) with $\alpha_4 < 0$ and $0 < \alpha_< < \alpha_> < 1$. On the right, the potential (\ref{total energy GM}) in the limit $\alpha_4 \rightarrow 0$ with $\alpha_6 > 0$.
\label{potentials picture}}
\end{figure}

\subsection{Explicit solution for $\alpha_4 <  0$}
\label{appendix solution alpha4 < 0}

\noindent Now we  consider of (\ref{total energy GM}) for $\alpha_4 <  0$.
When $\alpha_6^2-4\,\alpha_8\, \alpha_4> 0$ and $0<\alpha_< < \alpha_> <1$ (see the picture on the left in the figure \ref{potentials picture}), the solution reads\footnote{We choose the solution invariant under $y \rightarrow -\,y$.}
\begin{equation}
\label{solution u alpha4<0}
u(y)\,=\,\frac{\gamma}{\cos ( 2 \sqrt{|\alpha_4|}\,y)-R}
\end{equation}
with $\gamma$ and $R$ given by (\ref{gamma}) and (\ref{R}) respectively. Assuming $\alpha_6 >  0$, we have that $\alpha_8 <  0$ and $\alpha_4 <  0$ imply  $R>1$; therefore $u >0$ in (\ref{solution u alpha4<0}).
Notice that this solution oscillates at  $y \rightarrow \pm \infty$, therefore it does not satisfy the solitary wave boundary conditions. Moreover, the solution is such that $\alpha_< < u < \alpha_>$, therefore it remains far from $\theta=\pi$, namely we always have $u>0$.\\
Thus, because of the boundary conditions of our problem, this solution is irrelevant for our purposes.

\subsection{The limit $\alpha_4 \rightarrow 0$}
\label{appendix solution alpha4 = 0}

In this subsection we discuss the regime of $\alpha_4 \rightarrow 0$. In this limit the ratio $\Omega$ tends to one of two curves given in (\ref{bounds Omega}), i.e. the red and blue curves shown in the picture on the right of the figure \ref{picture potential GM} and also in the figure \ref{zoom bound omega picture}. In this limit $u=0$ is a triple zero of the potential in (\ref{total energy GM}).\\
Since $\gamma=2\alpha_4/|\alpha_6|(1+O(\alpha_4))$ and  $R=\alpha_6/|\alpha_6|(1+O(\alpha_4))$ for $\alpha_4 \rightarrow 0$, we have to distinguish between the case $R \rightarrow -1$ and $R \rightarrow 1$, which correspond to $\alpha_6<0$ and $\alpha_6>0$ respectively. 
From (\ref{alpha 6}), one observes that
\begin{equation}
\alpha_6>0
\hspace{1.5cm}
\Longleftrightarrow
\hspace{1.5cm}
\Omega\,<\,\frac{2(1-b)}{b}
\end{equation}
namely $\alpha_6<0$ above the horizontal dashed line in the figure \ref{zoom bound omega picture} (in the case of the conifold).
Considering the solutions (\ref{solution u}) and (\ref{solution u alpha4<0}), for every fixed $y$,  when $\alpha_6<0$ we have $u(y)=-\,\alpha_4/\alpha_6+O(\alpha_4)$, while when $\alpha_6>0$ we get
\begin{equation}
\label{small alpha4 alpha6>0}
 u(y)\,=\,\frac{\alpha_6}{\alpha_6^2\, y^2-\alpha_8} 
\end{equation}
up to $O(\alpha_4)$ terms.
One can check that (\ref{small alpha4 alpha6>0}) is the solution of (\ref{total energy GM}) with $\alpha_6>0$ and $\alpha_4=0$ (see the picture on the right in the figure \ref{potentials picture}) which vanishes at $|y| \rightarrow \infty$ with its derivatives (solitary wave solution).\\
As for the angular amplitude along the direction of $\theta$, we can take the limit $\alpha_4 \rightarrow 0^+$ of (\ref{Delta theta b}). For $\alpha_6<0$ we have $\Delta\theta_b =2\sqrt{-\,\alpha_4/\alpha_6}+O(\alpha_4^{3/2})$, while for $\alpha_6<0$ we find
\begin{equation}
\Delta\theta_b |_{\alpha_4=0}
\,=\,\pi-2\,\arctan \sqrt{-\,\frac{\alpha_8+\alpha_6}{\alpha_6}}
\,=\,\pi-2\,\arctan \sqrt{\frac{b\,\Omega - (1-b)}{2(1-b)- b\,\Omega}}
\phantom{xxxxx}
\end{equation}
which is finite.
Thus, when $\alpha_4=0$ and $\alpha_6>0$, we get a non trivial solution for $b<1$, which is present in particular on the conifold.\\
We remark that this feature does not occur for $S^3$ because in that case the potential $V(u)$ is cubic with a local maximum in $u=0$, therefore only the pointlike string limit can be performed.

\section{Useful integrals}
\label{appendix integrals}

In this appendix, we explicitly write the integrals which have been used to compute the dispersion relations.\\
Given the solution (\ref{solution u}) for $u(y)$, 
the definite integrals occurring in the finite combinations of charges are given by
\begin{eqnarray}
\label{U1}
U_1
& &\hspace{-.46cm}\equiv\,
\int_{-\infty}^{+\infty}u(y)\,dy
\;=\;
 \frac{2\,\gamma}{\sqrt{\alpha_4(1-R^2)}}
 \,\arctan \left(\sqrt{\frac{1+R}{1-R}}\;\right)
\\
\label{U2}
\rule{0pt}{.8cm}
U_2
& &\hspace{-.46cm}\equiv\,
\int_{-\infty}^{+\infty}u(y)^2\, dy
\;=\;
\frac{\gamma^2}{\sqrt{\alpha_4}\,(1-R^2)}
\left[\,
 \frac{2\,R}{\sqrt{1-R^2}}
 \,\arctan \left(\sqrt{\frac{1+R}{1-R}}\;\right) +\,1\,\right]
 \phantom{xxxx}
 \\
\label{Utilde}
\rule{0pt}{.9cm}
\widetilde{U}
& &\hspace{-.46cm}\equiv\,
\int_{-\infty}^{+\infty}\frac{u(y)}{1-u(y)}\, dy
\;=\;
 \frac{2\,\gamma}{\sqrt{\alpha_4(1-(R+\gamma)^2)}}
 \,\arctan \left(\sqrt{\frac{1+(R+\gamma)}{1-(R+\gamma)}}\;\right)\;.
 \phantom{xxxx}
\end{eqnarray}
As discussed in the subsection \ref{subsection explicit solution}, we have $|R|<1$ and $|R+\gamma|<1$, therefore these expressions are well defined. Moreover, from (\ref{U1}) and (\ref{U2}), it is straightforward to observe that
\begin{equation}
\label{U2 related to U1}
U_2\,=\,
-\,\frac{\alpha_6}{2\,\alpha_8}\,U_1-\,\frac{\sqrt{\alpha_4}}{\alpha_8}\;.
\end{equation}
Notice that in the limit $b \rightarrow 1$ (for which $R \rightarrow -1$, as discussed in the subsection \ref{subsection explicit solution}), $U_1$ and $U_2$ give finite expressions.

\section{The spin chain from the string sigma model}
\label{appendix spin chain}

In this appendix we perform the large spin limit introduced in \cite{Kruczenski:2003gt} for the background metric (\ref{metric b}), in order to study a possible spin chain related to the sigma model we are considering (see also \cite{Hernandez:2004uw}).\\
Given the metric (\ref{metric b}), we introduce the coordinate $\psi_1$ as follows \cite{Benvenuti:2005cz}
\begin{equation}
\label{psi1}
\psi \,=\,\frac{2}{b}\, t +\psi_1
\end{equation}
and we use the gauge choice $t = k\,\tau$. Then, we consider the following limit for the embedding fields \cite{Kruczenski:2003gt}\footnote{In (\ref{Kruczenski limit}), by $X^\mu$ we mean $(\psi_1,\theta,\phi)$.}
\begin{equation}
\label{Kruczenski limit}
k\,\rightarrow \,\infty
\hspace{2cm}
k\,\partial_\tau X^\mu \;\;\textrm{fixed}
\end{equation}
which implies that $\partial_\tau X^\mu \rightarrow 0$. 
In this limit the Virasoro constraints (\ref{VC1}) and (\ref{VC2}) with the background metric (\ref{metric b}) reduce respectively to\footnote{In (\ref{VC1 red}) we have employed (\ref{VC2 red}).}
\begin{eqnarray}
\label{VC1 red}
& & k\, (\partial_\tau \psi_1 - \cos \theta \,\partial_\tau \phi) 
+  \frac{1}{4}\big((\partial_\sigma \theta)^2+\sin^2\theta \,(\partial_\sigma \phi)^2\big)\;=\;0 \\
 \label{VC2 red}
\rule{0pt}{.5cm}
& & \partial_\sigma \psi_1 - \cos \theta \,\partial_\sigma \phi \;=\;0  
\end{eqnarray}
while for the Polyakov action (\ref{polyakov action}) we obtain
\begin{equation}
S\,= \,\frac{T}{2}\,\int d\tau \int_0^{2\pi}d\sigma\;
b 
\left[\,k\, (\partial_\tau \psi_1 - \cos \theta \,\partial_\tau \phi) 
-  \frac{1}{4}\big((\partial_\sigma \theta)^2+\sin^2\theta \,(\partial_\sigma \phi)^2\big)\,\right]\;.
\end{equation}  
Dropping the term containing the total derivative $\partial_\tau \psi_1$ and adopting $t = k\,\tau$ and $\tilde{\sigma} \equiv T\,b\,k\,\sigma $ as integration variables, we get
\begin{equation}
\label{polyakov action kruczenski limit}
S\,= \,- \,\frac{1}{2}\,\int dt\,\int_0^{\,b\,k\,\sqrt{\lambda}}d\tilde{\sigma}\,
\left[\,\cos \theta \,\partial_t \phi 
+  b^2 \frac{\lambda}{16\, \pi^2} 
\big((\partial_{\tilde{\sigma}} \theta)^2+\sin^2\theta \,(\partial_{\tilde{\sigma}} \phi)^2\big)\,\right]\;.
\end{equation}  
Now we want to recover this action as the continuum limit of a discrete sigma model. \\
Let us consider the $SU(2)$ spin $1/2$ chain with $L$ sites. Applying an arbitrary $SU(2)$ rotation to the maximally polarized state at each site, we get the following normalized state
\begin{equation}
\label{state}
|  \hat{n} \rangle \,=\,
\cos (\theta/2)  \,e^{i\,\phi/2}\, |  1 \rangle\,+\,
\sin (\theta/2)  \,e^{-i\,\phi/2} \,|  2 \rangle
\end{equation}  
where the continuum variables $\theta \in [0,\pi]$ and $\phi \in [0,2\pi]$ depend on $(t,\tilde{\sigma})$.
The scalar product for two coherent states of this form reads
\begin{equation}
\label{scalar product}
\langle \hat{n}'  |  \hat{n} \rangle \,=\,
\cos \big((\theta-\theta')/2\big) \cos \big((\phi-\phi')/2\big)
+ i  \cos \big((\theta+\theta')/2\big) \sin \big((\phi-\phi')/2\big)\;.
\end{equation}  
The action of the discrete sigma model is
\begin{equation}
\label{action discrete sigma model}
S\,=\,  \sum_{l\,=\,1}^L \int dt
\left[\, i \,\frac{d}{d\rho}\, \langle \hat{n}_l(t) |  \hat{n}_l(t+\rho) \rangle|_{\rho=0}
-\,
\langle \hat{n}_l(t), \hat{n}_{l+1}(t) |  H_{l,l+1} |\hat{n}_l(t), \hat{n}_{l+1}(t) \rangle
\,\right]
\end{equation}  
where the first term is the Wess-Zumino term and for the nearest neighbour Hamiltonian we take
\begin{equation}
\label{hamiltonian spin chain b}
H_{l,l+1}
\,=\,b^2\, \frac{\lambda}{8\pi^2}  \,\big(1 - P_{l,l+1}\big)\;.
\end{equation}  
The operator $P_{l,l+1}$ is the exchange operator between the sites $l$ and $l+1$: given a state $|  \alpha \rangle$ at the site $l$ and a state $|  \beta \rangle$ at the site $l+1$, one has 
$\langle \alpha, \beta |  P_{l,l+1} |\alpha, \beta \rangle = |\langle \alpha  |  \beta \rangle|^2$.\\
Now we take the continuum limit of the discrete sigma model, namely we send the number of sites $L \rightarrow \infty$ and consider slowly varying spin configurations only, which allows to replace finite differences between the spin variables at neighbouring sites by derivatives.
The distance $2\pi/L$ between two adjacent sites goes to zero and the hamiltonian term turns out to be $O((2\pi/L)^2)$.
Introducing the variable $\sigma\in [0,2\pi]$, for the continuum limit of (\ref{action discrete sigma model}) we obtain
\begin{equation}
\label{action discrete sigma model limit}
S\,=\,  - \,\frac{L}{4\pi}\, \int dt \int_0^{2\pi}d \sigma\;
\left[\,
\cos \theta \,\partial_t \phi 
+  b^2 \frac{\lambda}{4 L^2}\,
\big((\partial_{\sigma} \theta)^2+\sin^2\theta \,(\partial_{\sigma} \phi)^2\big)\
\,\right]
\end{equation}  
which is exactly the expression (\ref{polyakov action kruczenski limit}) obtained from the Polyakov action, once we identify $L=b\,k\,\sqrt{\lambda}$ and rescale the spatial variable according to the previous definition of $\tilde{\sigma}$.\\
With respect to the case of the $SU(2)$ spin chain found from the string sigma model on $\mathbb{R} \times S^3$ \cite{Kruczenski:2003gt}, the Hamiltonian (\ref{hamiltonian spin chain b}) has a global factor $b^2$. Notice that also in the BMN limit considered in the subsection \ref{BMN and GKP} (see (\ref{GM dispersion in BMN limit})) we find the same factor $b^2$ in front of $\Delta\varphi^2$, that we would like to interpret as $p^2$ of the spin chain magnon.
Thus, it seems that we can give a spin chain interpretation of the BMN regime of our giant magnon through the Hamiltonian $H= \sum _{l=1}^L H_{l,l+1}$ with $H_{l,l+1}$ given by (\ref{hamiltonian spin chain b}). We remark that the relation of the dilatation operator of the $\mathcal{N}=1$ gauge theory of \cite{Klebanov:1998hh} with the Hamiltonian of some spin chain, if it exists, is still unknown.

%%%%%%%%%%%%%%%%%%%%%%%%%%%%%%%%%%%%%%%%%%%%%%%%%%%

\end{document}